\documentclass[onecolumn]{aa}  
\usepackage{graphicx}
\usepackage{txfonts}
\usepackage[dvipsnames]{xcolor}
\usepackage[colorlinks=true,linkcolor=NavyBlue,citecolor=NavyBlue]{hyperref}

\begin{document}

   \title{Identifying Footpoints of Pre-eruptive and Coronal Mass Ejection Flux Ropes with Sunspot Scars}

   \author{Chen Xing \inst{1,2}\fnmsep\thanks{Corresponding author},
               Guillaume Aulanier \inst{1,3},
               Brigitte Schmieder \inst{4,5,6},
               Xin Cheng \inst{2}, \and
               Mingde Ding \inst{2}
               }

   \institute{Sorbonne Université, Observatoire de Paris-PSL, École Polytechnique, IP Paris, CNRS, Laboratoire de Physique des Plasmas, Paris, France\\
                  \email{chenxing@smail.nju.edu.cn}
                  \and
                  School of Astronomy and Space Science, Nanjing University, Nanjing, China
                  \and
                  Rosseland Centre for Solar Physics (RoCS), Institute of Theoretical Astrophysics, Universitetet i Olso, Oslo, Norway
                  \and
                  LESIA, Observatoire de Paris, Université PSL, CNRS, Sorbonne Université, Université de Paris, 5 place Jules Janssen, 92190 Meudon, France
                  \and
                  Centre for Mathematical Plasma Astrophysics, Dept. of Mathematics, KU Leuven, 3001 Leuven, Belgium
                  \and
                  SUPA, School of Physics \& Astronomy, University of Glasgow, G12 8QQ, UK
                   }

   \date{Received ...; accepted ...}

 
  \abstract
   {The properties of pre-eruptive structures and coronal mass ejections (CMEs) are characterized by those of their footpoints, the latter of which thus attract great interest. However, how to identify the footpoints of pre-eruptive structures and how to identify the footpoints with ground-based instruments, still remain elusive.}
   {In this work, we study an arc-shaped structure intruding in the sunspot umbra. It is located close to the (pre-)eruptive flux rope footpoint and is thus expected to help identify the footpoint.}
   {We analyse this arc-shaped structure, which we name as ``sunspot scar'', in a CME event on 2012 July 12 and in two CME events in observationally-inspired MHD simulations performed by OHM and MPI-AMRVAC.}
   {The sunspot scar has a more inclined magnetic field with a weaker vertical component and a stronger horizontal component relative to that in the surrounding umbra and manifests as a light bridge in the white light passband. The hot field lines anchored in the sunspot scar are spatially at the transition between the flux rope and the background coronal loops, and temporally in the process of the slipping reconnection which builds up the flux rope.}
   {The sunspot scar and its related light bridge mark the edge of the CME flux rope footpoint, and especially, the edge of the pre-eruptive flux rope footpoint in the framework of ``pre-eruptive structures being flux ropes''. Therefore, they provide a new perspective for the identification of pre-eruptive and CME flux rope footpoints, and also new methods for studying the properties and evolution of pre-eruptive structures and CMEs with photospheric observations only.}


   \titlerunning{Identifying Footpoints of Pre-eruptive and CME Flux Ropes with Sunspot Scars}
   \authorrunning{Xing et al.}
     
   \maketitle
%

\nolinenumbers

\section{Introduction}\label{sec1}
The Sun frequently releases rapid ejections of plasmas, known as coronal mass ejections (CMEs), into the interplanetary space. When CMEs propagate close to the Earth and interact with the Earth's magnetosphere, they could induce damages to human high-tech activities such as satellite communications and power transmission \citep{Elovaara2007}. The forecast of CMEs, which is important to prevent these space hazards, relies on a thorough understanding of the properties of CMEs. Generally, it is considered that a flux rope, in which magnetic field lines intertwine with each other, is the main magnetic configuration of the CME \citep{Chen1997,Dere1999}; it is even likely that of the pre-eruptive structure of many CMEs (see the reviews by \citeauthor{Cheng2017}\ \citeyear{Cheng2017} and \citeauthor{Patsourakos2020}\ \citeyear{Patsourakos2020}) as evidenced by the filament \citep{Aulanier1998,Guo2010,Schmieder2013} and the hot channel \citep{Cheng2012,Zhang2012}.

The footpoints of flux ropes, taking advantage of their measurable magnetic field and close links to pre-eruptive structures and CMEs, make great contributions to understanding the properties and evolution of these (pre-)eruptive structures. For example, with the magnetic field in the identified footpoints, \cite{Wang2019,Wang2017} showed the increase of twist in pre-eruptive and CME flux ropes, which demonstrates the building up of these structures. During the eruption, the drifting and deformation of CME footpoints could reveal the magnetic reconnection processes related to the CME evolution \citep{Aulanier2019,Gou2023}, and these processes are further supported by the evolution of the toroidal flux in the CME footpoint \citep{Xing2020}. Furthermore, before CMEs pass the Alfvénic surface in the solar wind, the conditions of CME footpoints always have effects on the shape and extension of CMEs, as the magnetic tension force can still propagate all along CMEs then. Lastly, the location and shape of footpoints and the magnetic field in footpoints are also essential input parameters in many extrapolation method for reconstructing the coronal magnetic field in the pre-eruptive flux rope, e.g., the flux rope insertion method \citep{Su2011} and the Regularized Biot-Savart Law method \citep{Titov2021,Guo2023}. An accurate identification of the flux rope footpoints, as the basis of these works, is thus extremely important.

So far, two methods have been proposed to identify CME footpoints in observations. First, the two footpoints of the CME flux rope are considered to correspond to a pair of core dimming regions where the EUV intensity is significantly reduced during the eruption \citep{Sterling1997,Cheng2016,Wang2017}. The core dimming is believed to be caused by the plasma rarefaction in the CME foot as a result of the flux rope eruption/expansion \citep{Harra2001,Tian2012} or to be caused by the temporary generation of a transient solar wind at the footpoints of the CME expanding field lines \citep{Lorincik2021}. Second, the CME footpoint is usually identified as the region (partially) enclosed by the hooked part of the flare ribbon \citep{Wang2017,Aulanier2019}, as the footprint of the quasi-separatix layers (QSLs; \citeauthor{Priest1995}\ \citeyear{Priest1995}) matches well with both the flare ribbon hook \citep{Savcheva2015,Zhao2016} and the boundary of the flux rope footpoint \citep{Demoulin1996,Janvier2014}. In addition, some recent works also identify the footpoint with a combination of these two features (e.g., \citeauthor{Xing2020}\ \citeyear{Xing2020}). However, we note that, on the one hand, these two methods are usually only applicable to CMEs rather than pre-eruptive structures, as the core dimming and the flare ribbon are accompanying phenomena of the eruption; on the other hand, the pre-eruptive flux rope footpoint is always not the same as the CME footpoint, as the footpoint is built up under the magnetic reconnection among overlying field lines \citep{Aulanier2019} and deformed under the magnetic reconnection in the flux rope and that between the flux rope and the overlying field \citep{Aulanier2019,Gou2023} during the eruption. Therefore, it is still elusive to identify the footpoints of pre-eruptive structures of CMEs.

The recent studies showed that the CME evolution in the corona could influence the photospheric magnetic field, that, the so called "the tail wagging the dog" phenomenon \citep{Aulanier2016}. For example, the photospheric magnetic field is found to become more horizontal close to the polarity inversion line (PIL) of the CME source region \citep{Sun2012,Wang2012a} while more vertical in the periphery of the source region \citep{Liu2005,Wang2012b} after the eruption. The change in the magnetic field may result from that the energy released during the eruption induces a reduction of the magnetic pressure in the core of the active region \citep{Wang2012b}. Recently, \cite{Barczynski2019} gave another explanation to these changes, that the more horizontal field close to the PIL is due to the reconnection-driven contraction of post-flare loops and the more vertical field in the periphery is a result of CME-driven stretching of coronal loops. A similar idea is also proposed by \cite{ChengX2016} that the stretching of CME legs leads to the photospheric magnetic field becoming more vertical. The influence of the CME evolution on the photosphere even manifests in the white light passband, showing as the penumbra darkened near the PIL while weakened in the outer periphery of sunspots \citep{Liu2005,Wang2012b}. In addition, \cite{Wang2021} showed that the magnetic reconnection during the erosion-stage of the CME eruption could lead to an enhancement of both the vertical and horizontal components of the photospheric magnetic field in the ribbon-swept region. We note that all of the above results suggest that the features of the photospheric magnetic field in the CME source regions could reflect the properties and evolution of CMEs.

In this work, we demonstrate such a phenomenon in both observations and simulations, that an arc-shaped structure intruding in the sunspot umbra marks the footpoint edge of the pre-eruptive flux rope of the CME. The simulations even suggest that this structure marks the footpoint edge for the flux rope during the eruption. We highlight that this is a new perspective for identifying the flux rope footpoints, applicable to both the ground-based and space-based telescopes. In the following, we show this phenomenon in observations in Section 2 and that in simulations in Section 3. We also demonstrate the nature of this phenomenon in Section 3, after which a summary is presented in Section 4 and several discussions are given in Section 5.

\section{Sunspot Scar in a CME/flare Event on 2012 July 12}\label{sec2}
\subsection{Data}\label{sec21}
In this section, we study this arc-shaped structure in a CME/flare event on 2012 July 12. The evolution of this event is exhibited by EUV images at 131 \AA\ and 304 \AA\ passbands, which are observed by Atmospheric Imaging Assembly (AIA; \citeauthor{Lemen2012}\ \citeyear{Lemen2012}) on board \textit{Solar Dynamics Observatory} (\textit{SDO}; \citeauthor{Pesnell2012}\ \citeyear{Pesnell2012}), and also SHARP CEA-coordinate continuum intensity maps, SHARP CEA-coordinate vector magnetic field maps and helioprojective-cartesian-coordinate vector magnetic field images which are observed by Helioseismic and Magnetic Imager (HMI; \citeauthor{Scherrer2012}\ \citeyear{Scherrer2012}) on board \textit{SDO}. In this work, we use EUV images with a cadence of 12 seconds, and magnetic field maps and continuum intensity maps with a cadence of 12 minutes.

\subsection{Overview of the CME/Flare Event}\label{sec22}
\begin{figure*}
\centering
\includegraphics[width=\hsize]{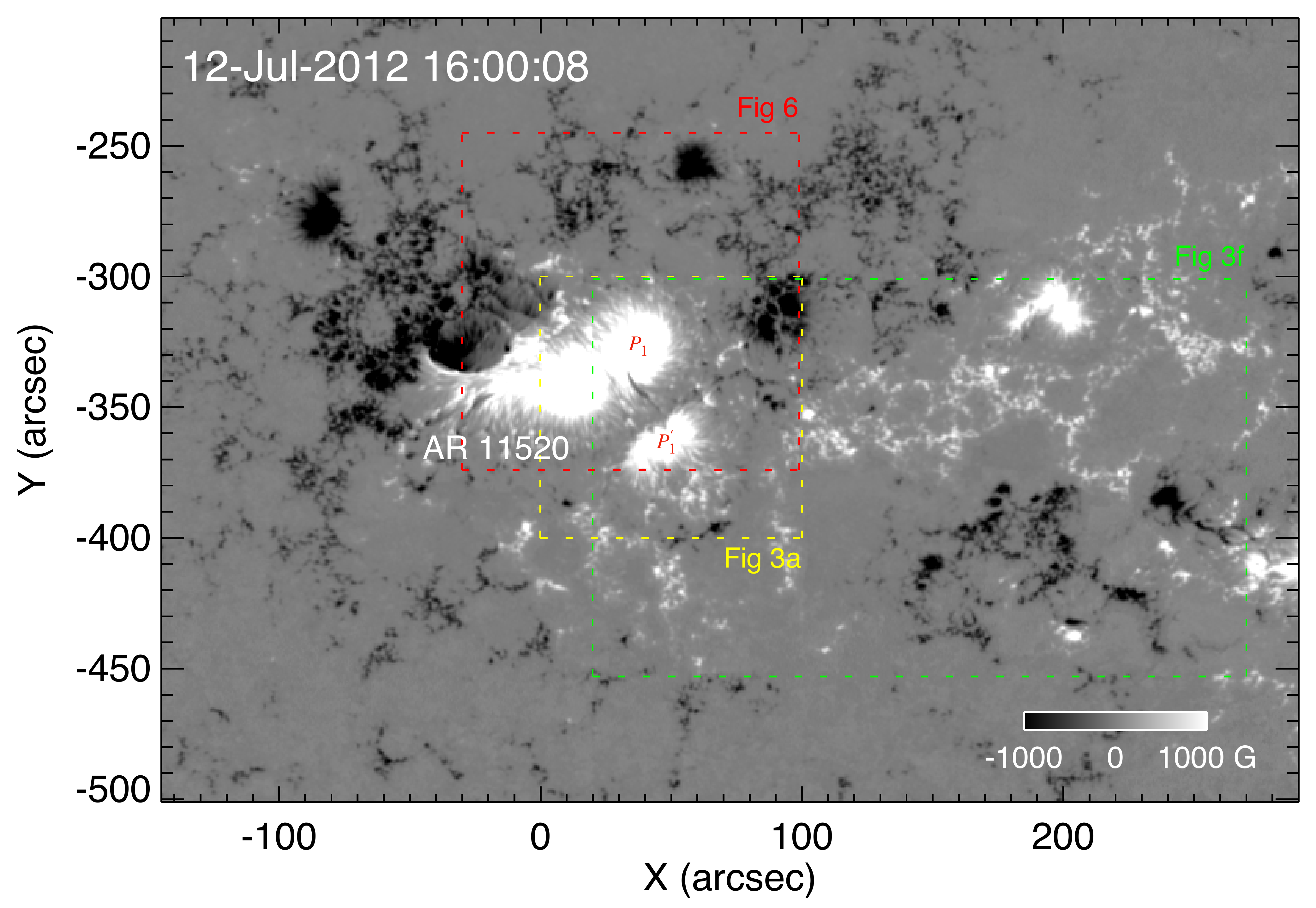}
\caption{Vertical magnetic field ($B_z$, component of magnetic field perpendicular to the solar surface) image of the source region of the CME/flare event on 2012 July 12. The field of view is the same as that in Fig.~\ref{fig2}. The yellow, green, and red dashed boxes show the field of view in Fig.~\ref{fig3}a, Fig.~\ref{fig3}f, and Fig.~\ref{fig6}, respectively. $P_1$ and $P_1^{'}$ mark the main positive polarity and the adjacent positive polarity, respectively.}
\label{fig1}
\end{figure*}

A major CME and its associated X1.4 class flare occurred on 2012 July 12 in the active region NOAA 11520 (see Fig.~\ref{fig1}). The GOES soft X-ray flux rapidly increases since $\sim$16:10 UT and peaks at $\sim$16:49 UT. This event is thoroughly studied by many works, e.g., that focusing on the slipping reconnection related to the formation of the (pre-)eruptive structure \citep{Dudik2014} and that studying the formation of a double-decker flux rope in the source region \citep{Cheng2014}. In the following we briefly review the evolution of this event, and we refer readers to these previous works for more details.

\begin{figure*}
\centering
\includegraphics[width=\hsize]{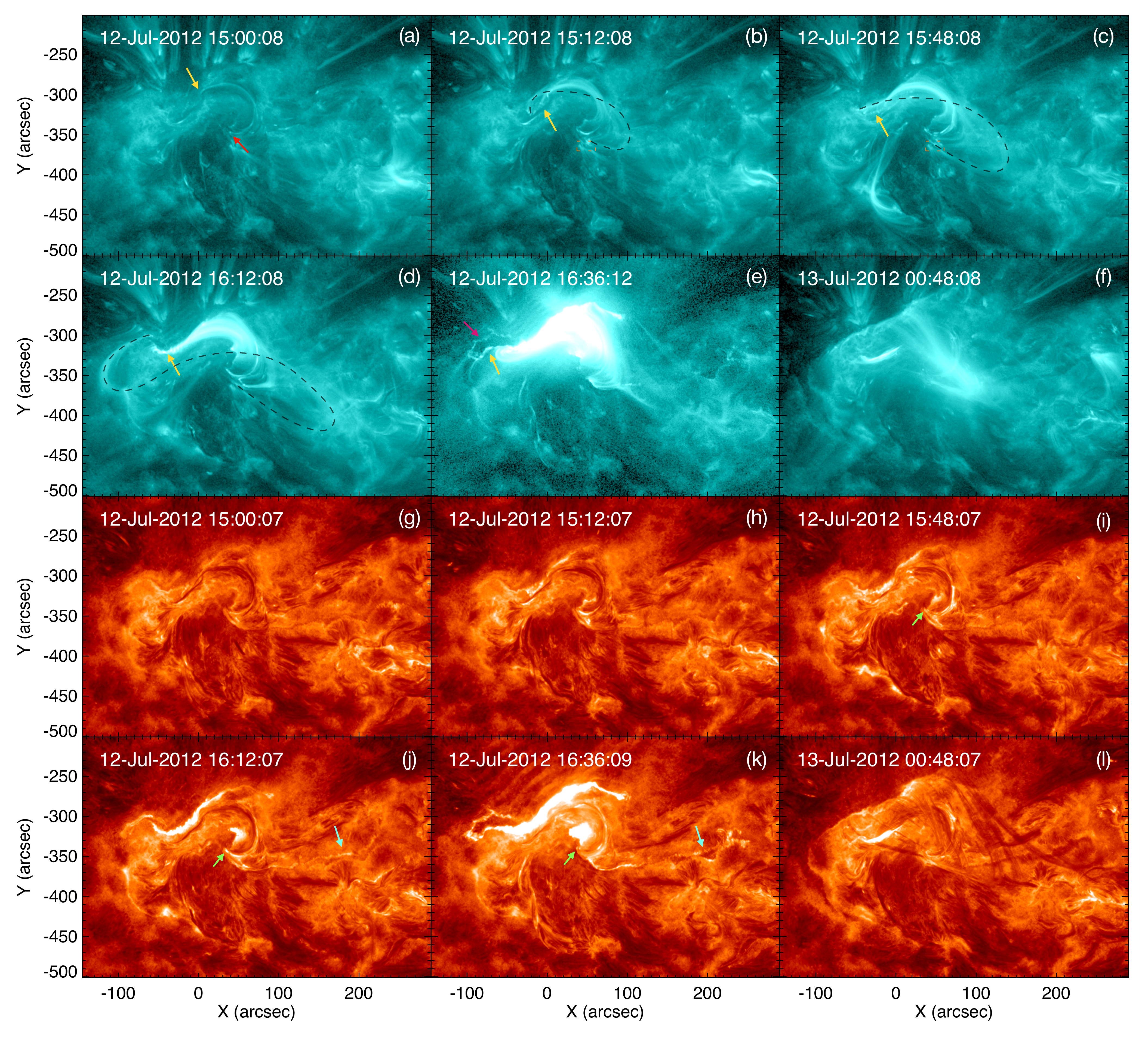}
\caption{AIA 131 \AA\ images (panels a-f) and 304 \AA\ images (panels g-l) of the source region of the CME/flare event on 2012 July 12. The red arrow in panel a marks the first bright arcade at 15:00 UT. The yellow arrows in panels a-e mark the slipping motion of the eastern footpoints of the bright arcades in the negative polarity. The dashed lines in panels b-d show the pre-eruptive hot channel. The orange dashed box in panels b and c marks the western footpoint region of the hot channel. The eastern footpoint of the CME is marked by the purple arrow in panel e and the western one is marked by the blue arrow in panel k. The blue arrow in panel j shows the narrow ribbon lane along which the bright loops slip. The green arrows in panels i-k point the western footpoint of the low-lying filament.}
\label{fig2}
\end{figure*}

As shown in Fig.~\ref{fig2}a-f and Movie 1, the formation of the pre-eruptive structure starts at $\sim$15:00 UT, represented by the appearance of the first bright arcade (marked by the red arrow in Fig.~\ref{fig2}a). Afterwards, several bright arcades rapidly develop: their eastern footpoints (pointed by yellow arrows in Fig.~\ref{fig2}b,c) quickly slip towards east, starting from the eastern footpoint of the first bright arcade (marked by the yellow arrow in Fig.~\ref{fig2}a) and along the bright ribbons in the negative polarity (visible at 304 \AA\ passband; see Fig.~\ref{fig2}h,i). The details of this process are also exhibited in Fig. 6 and Fig. 8 in \cite{Dudik2014}. The pre-eruptive structure, appearing as a hot channel at 131 \AA\ passband (marked by black dashed lines in Fig.~\ref{fig2}b,c), is gradually formed at the same time. In this process, the fainter pre-eruptive structure lies above the bright arcades, with its eastern footpoint rooted close to the easternmost footpoints of bright arcades and also slipping along the bright ribbons in the negative polarity. The western footpoint of the hot channel (marked by the orange dashed box in Fig.~\ref{fig2}b-c and Fig.~\ref{fig3}a-d) is mainly anchored in a small adjacent positive sunspot ($P_1^{'}$; see Fig.~\ref{fig3}, around $(X,Y)=(50,-365)$ arcsecs), close to those of the bright arcades in the main positive sunspot ($P_1$; also see Fig.~\ref{fig3}, around $(X,Y)=(40,-335)$ arcsecs). The eruption of the hot channel starts at around 16:12 UT (Fig.~\ref{fig2}d), and the slipping motion in the negative polarity continues at the onset of and during the eruption (marked by yellow arrows in Fig.~\ref{fig2}d,e). The eastern ribbon hook (pointed by the purple arrow in Fig.~\ref{fig2}e) represents the eastern CME footpoint and develops rapidly during the eruption.

At AIA 304 \AA\ passband (see Fig.~\ref{fig2}g-i and Movie 2), an undisturbed filament appears in this active region during the above whole process. It lies below the pre-eruptive hot channel before the eruption (Fig.~\ref{fig2}h,i), while it still stays on the spot and below the post-flare loops during and after the eruption (Fig.~\ref{fig2}j-l). This indicates that the eruptive hot channel and the filament are in the upper part and the lower part of a double-decker structure, respectively, in agreement with the views in \cite{Cheng2014}. The filament is quite stable and inactive before and during the eruption, unaffected by the formation and eruption of the hot channel.

\subsection{Evolution of the Western Leg of the Hot Channel}\label{sec23}
\begin{figure*}
\centering
\includegraphics[width=0.9\hsize]{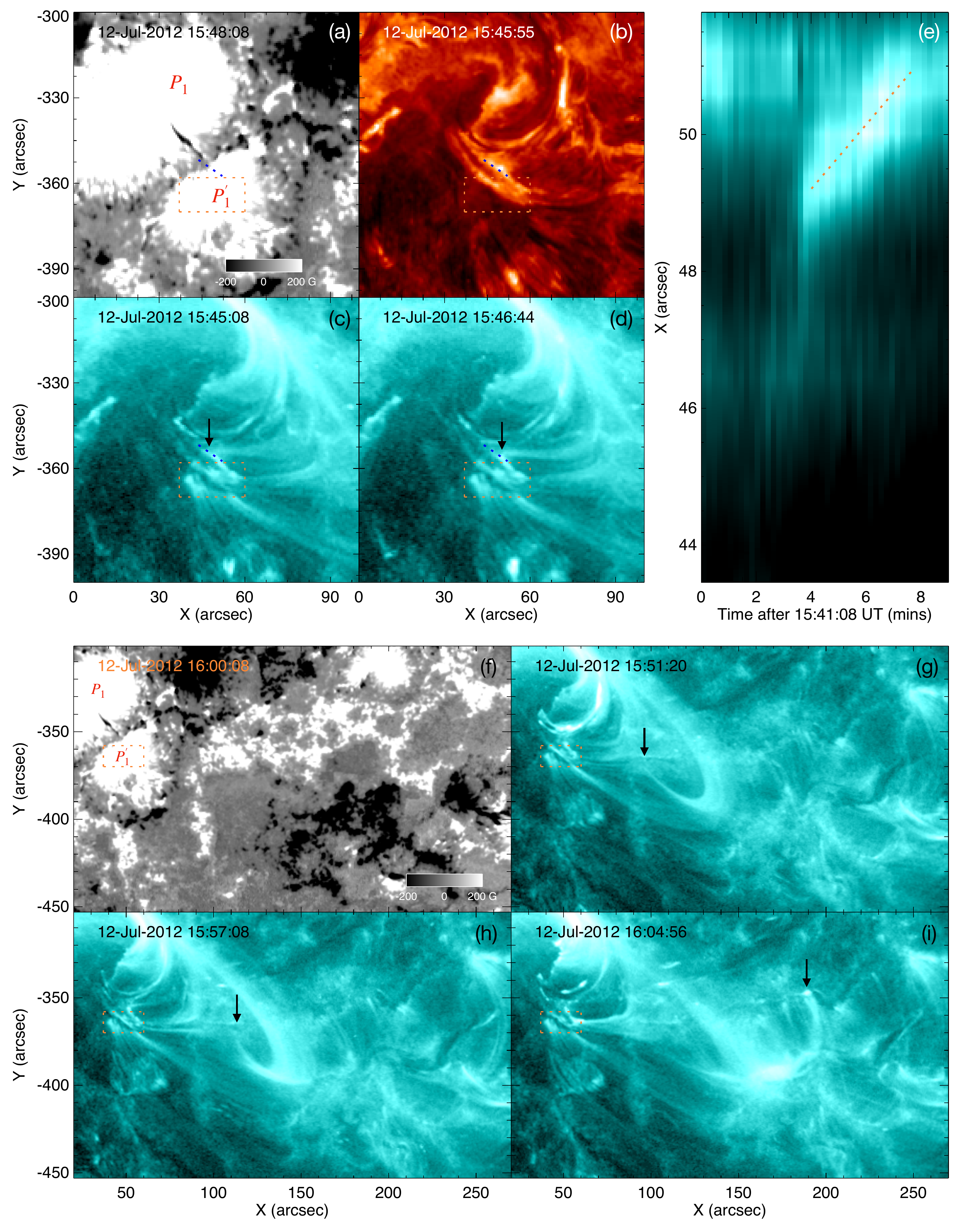}
\caption{Slipping motions of loops in the western leg of the hot channel, from the main positive sunspot to the adjacent positive sunspot (panels a-e) and further from the adjacent positive sunspot towards west (panels f-i). (a) Vertical magnetic field image showing the main positive sunspot ($P_1$) and the adjacent positive sunspot ($P_1^{'}$). (b) 304 \AA\ image with a same field of view as panel a. (c)-(d) 131 \AA\ images with a same field of view as panel a. In panels a-d, the orange dashed box marks the western footpoint region of the hot channel, and the blue dashed line represents the bright lane along which the bright loops at 131 \AA\ passband slip. The black arrows mark the footpoints of the slipping bright loops. (e) Time-slice plot of 131 \AA\ intensity along the blue dashed line in panels c-d. The orange dashed line in it marks the slipping motion of the bright loop footpoint. (f) Vertical magnetic field image showing the main positive polarity ($P_1$), the adjacent positive polarity ($P_1^{'}$) and the western positive-polarity faculae. (g)-(i) 131 \AA\ images with a same field of view as panel f. The orange dashed box in panels f-i is the same as that in panels a-d. The black arrows mark the footpoints of the slipping bright loops.}
\label{fig3}
\end{figure*}

Especially, we focus on the evolution of the western leg of the hot channel. Before the eruption onset, as shown in Fig.~\ref{fig3}a-d, some bright loops (see their footpoints pointed by black arrows in Fig.~\ref{fig3}c,d) slip along a bright lane (visible at 131 \AA\ and 304 \AA\ passbands; marked by the blue dashed line in Fig.~\ref{fig3}a-d), from the western foot of the filament (in the main positive sunspot) to the western foot of the hot channel (in the adjacent positive sunspot). This slipping motion is also demonstrated in the time-slice plot of the 131 \AA\ intensity along the blue dashed line (Fig.~\ref{fig3}e). These suggest that the hot channel loops are originally rooted close to the filament footpoint, while they soon slip to the hot channel footpoint (the orange dashed box) that we see.

Starting at $\sim$15:51 UT (still before the eruption), some loops in the western leg of the hot channel further slip towards west (see their footpoints pointed by black arrows in Fig.~\ref{fig3}g-i) along a narrow extended ribbon (pointed by the blue arrow in Fig.~\ref{fig2}j) in the western positive-polarity faculae (with X-coordinate in the range of [100,250] arcsecs), while the rest part of the western leg is still anchored in the adjacent positive sunspot. After the eruption onset, the bright loops continue to slip along the extended ribbon, the latter of which soon develops into the western ribbon hook (pointed by the blue arrow in Fig.~\ref{fig2}k) and represents the western CME footpoint. It is ambiguous if previous hot channel footpoints in the adjacent positive sunspot are a part of the CME footpoint, as the field lines anchored there are faint during the eruption.

\subsection{Properties of the Sunspot Scar}\label{sec24}

\begin{figure*}
\centering
\includegraphics[width=\hsize]{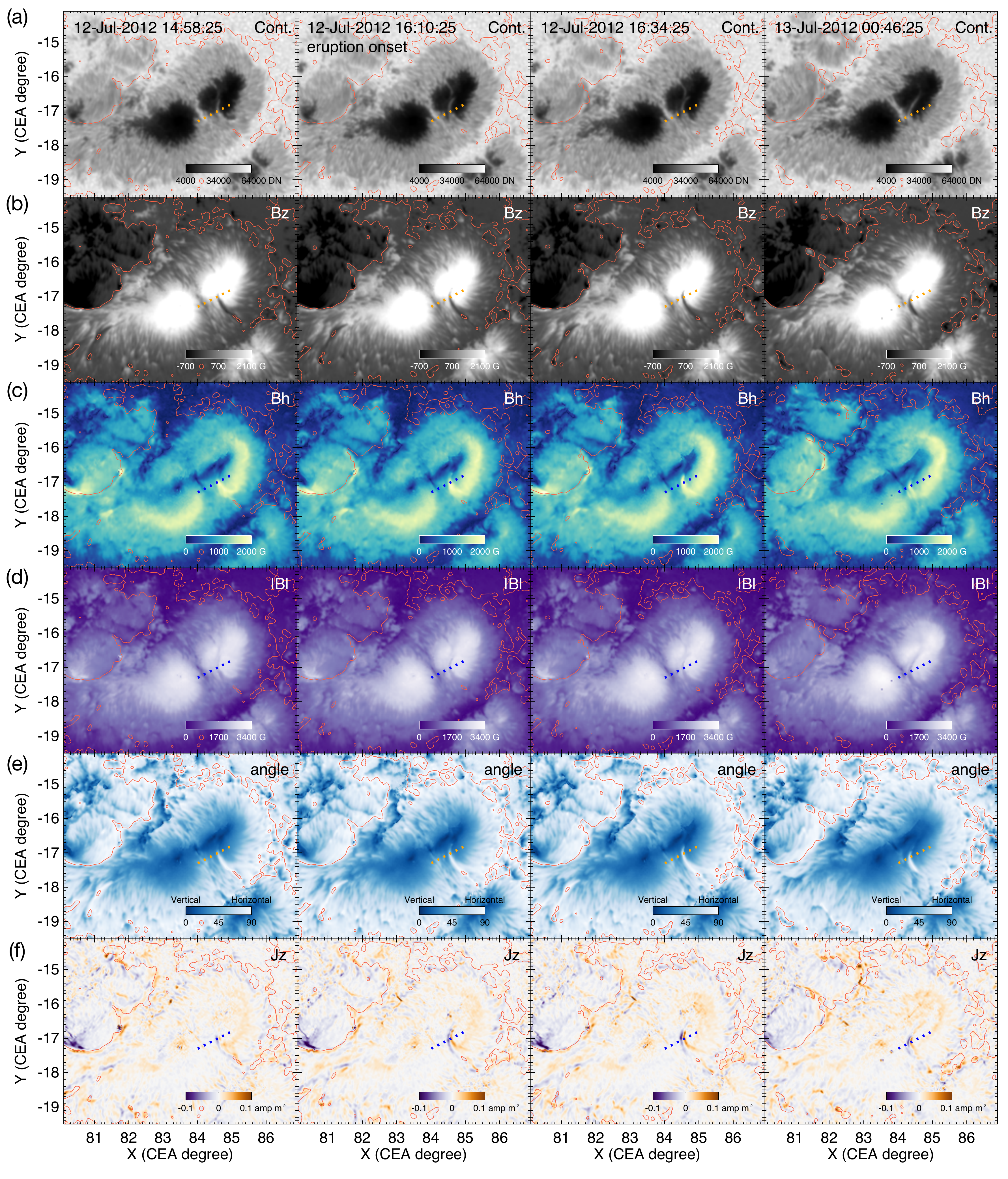}
\caption{Continuum intensity (Cont.; panel a), vertical magnetic field strength ($B_z$; panel b), horizontal magnetic field strength ($B_h$; panel c), total magnetic field strength ($|B|$; panel e), inclination angle of the direction of magnetic field to the vertical direction (angle; panel e), and the vertical current density ($J_z$; panel f) images of the source region, which are in SHARP CEA-coordinate and at four moments. The time of images in each column is shown in the first row of them. The red contour in each sub-panel shows the PIL. The orange or blue dashed line in each sub-panel is the slit crossing though the sunspot scar, along which the 1D parameters are shown in Fig.~\ref{fig5}.}
\label{fig4}
\end{figure*}

In the following, we study the source region of the CME/flare event in detail at four moments, i.e., 14:58 UT, 16:10 UT, 16:34 UT, and 00:46 UT on July 13, which represent the stage before, at the onset of, during, and after the eruption, respectively. As shown in Fig.~\ref{fig4}a,b, both the main sunspot umbra in the source region and its corresponding positive polarity (around $(X,Y)=(84,-17)$ CEA degrees) are composed of two parts.

\begin{figure*}
\centering
\includegraphics[width=\hsize]{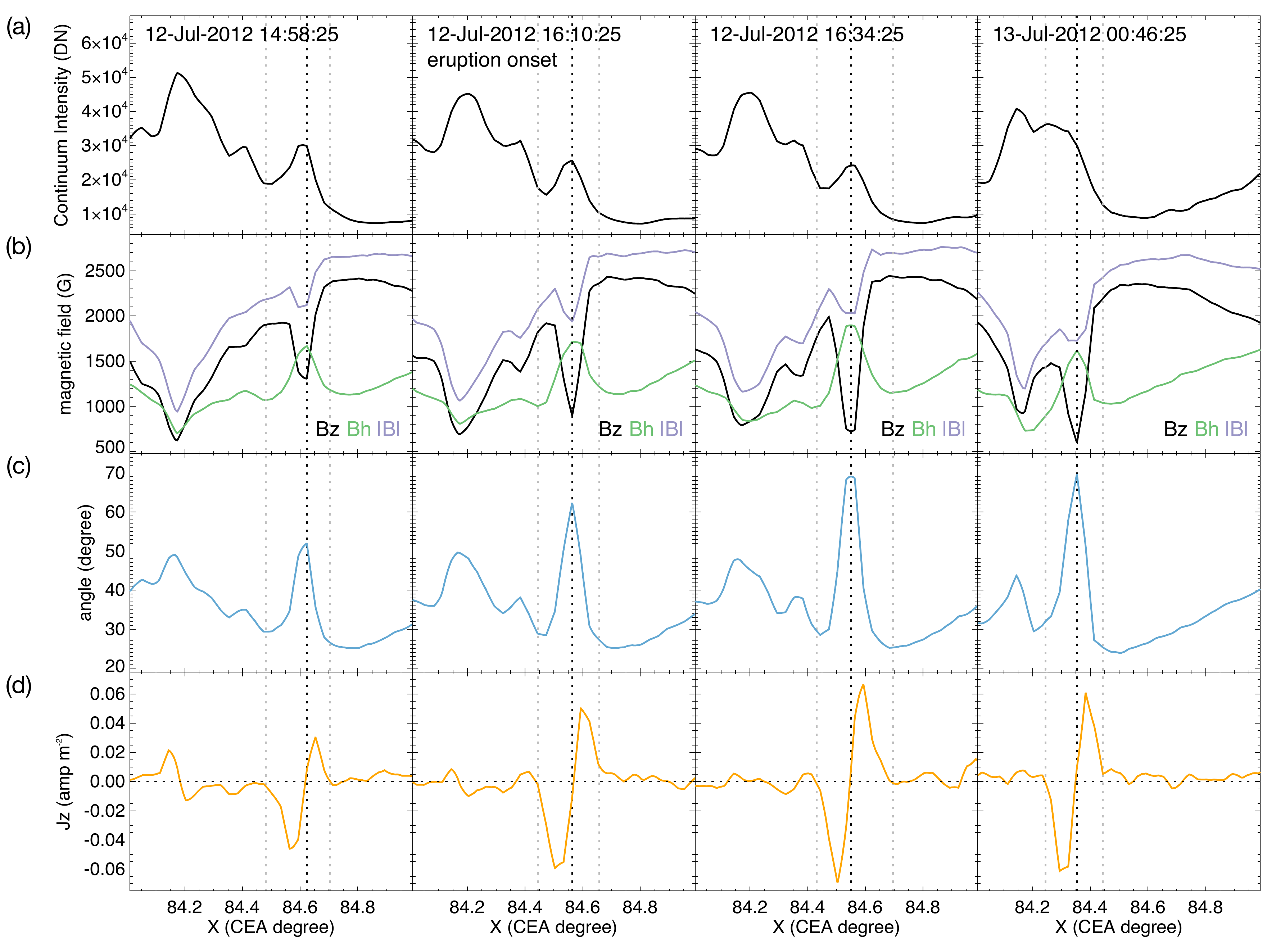}
\caption{The 1D parameters along the slit crossing through the sunspot scar (dashed lines in Fig.~\ref{fig4}). From the top to the bottom, they are continuum intensity (panel a), total magnetic field strength ($|B|$; panel b), vertical magnetic field strength ($B_z$; panel b), horizontal magnetic field strength ($B_h$; panel b), inclination angle (panel c), and vertical current density ($J_z$; panel d). The time of sub-panels in each column is shown in the first row of them. The black vertical dashed line in each column marks the dip of $B_z$ in the sunspot scar, which is regarded as the sunspot scar center. The grey dashed lines in each column show the range of the two peaks in $J_z$ which are at two sides of the sunspot scar, representing the edges of the sunspot scar. The black horizontal dashed line in panel d marks the value where the current density equals zero.}
\label{fig5}
\end{figure*}

Since $\sim$13:34 UT, an arc-shaped structure wedges into the western part of the main positive sunspot (Fig.~\ref{fig4}b and Movie 3), which we name as ``sunspot scar" hereafter throughout the rest of this paper. The sunspot scar appears and continuously drifts towards the south-east both before and after the eruption onset. It shows a significant decrease in the vertical magnetic field strength ($B_z$; component of magnetic field perpendicular to the solar surface) as compared to those in the surrounding umbra, while its polarity still remains positive almost everywhere. In Fig.~\ref{fig5}b, we shows $B_z$ along the orange slit crossing through the sunspot scar (see the slit in Fig.~\ref{fig4}b): $B_z$ is reduced by 39\%-67\% at the center of the sunspot scar at four moments, compared to the average of $B_z$ at two edges of the sunspot scar which are marked by the grey dashed lines in Fig.~\ref{fig5}. In the white light passband, the sunspot scar appears as a curved lane of bright materials, that resembles a narrow light bridge, in the western part of the umbra (Fig.~\ref{fig4}a and Movie 3). The sunspot scar matches well with this narrow light bridge at the first three moments, as the local dip of $B_z$ in the scar (marked by the black dashed line in Fig.~\ref{fig5}b) is co-spatial with the local peak of the continuum intensity in the light bridge (see Fig.~\ref{fig5}a). For the fourth moment, we consider that they still match with each other; the mismatch between the local peak of continuum intensity and the local dip of $B_z$ (Fig.~\ref{fig5}a,b) is due to that the curved drifting scar-related light bridge merges with the other brighter light bridge which separates the two parts of the umbra (see Fig.~\ref{fig4} and Movie 3). At the center of the sunspot scar, the continuum intensity increases by 72\%-95\% at the first three moments (see that in Fig.~\ref{fig5}a which is along the orange slit in Fig.~\ref{fig4}a).

Furthermore, the sunspot scar also shows marked features in aspect of other parameters of magnetic field at all four moments (see Fig.~\ref{fig4}c-f). The details of these parameters, sampled by those along the slit crossing through the sunspot scar (the orange and blue dashed lines in Fig.~\ref{fig4}c-f), are exhibited in Fig.~\ref{fig5}b-d. In the sunspot scar, there is an obvious enhancement in the horizontal magnetic field strength ($B_h$; component of magnetic field parallel to the solar surface; Fig.~\ref{fig4}c and Fig.~\ref{fig5}b); along the slit, $B_h$ is enhanced by 48\%-76\% at the sunspot scar center at four moments, compared to the average of $B_h$ at two edges of the sunspot scar. The reduction in $B_z$ and the enhancement of $B_h$ together lead to a more horizontal field, characterized by the larger inclination angle between the magnetic field direction and the vertical direction (Fig.~\ref{fig4}e and Fig.~\ref{fig5}c). At the center of the sunspot scar, the inclination angle is increased by 24-42 degrees at four moments. The total magnetic field strength ($|B|$) in the sunspot scar is a bit weaker than those in the surrounding umbra (Fig.~\ref{fig4}d), with a 12\%-18\% reduction at the scar center at four moments (see Fig.~\ref{fig5}b). The vertical current density ($J_z$) is close to zero at the center of the scar (see Fig.~\ref{fig4}f and Fig.~\ref{fig5}d), however, it has a positive peak on one side of the sunspot scar which the scar curves inwards to (hereafter referred to as \textit{inner side}) and a negative peak on the other side (hereafter referred to as \textit{outer side}).

We note that the sunspot scar is significantly different from the penumbra, as the above parameters, especially the inclination angle and $B_z$, are homogeneous in the most part of the former while heterogeneous in the latter. However, we also notice that sometimes the sunspot scar exhibits opposite $B_z$ in a small area at its end close to the penumbra (contoured by the red contours around $(X,Y)=(85,-17.6)$ CEA degrees in $B_z$ maps at first three moments in Fig.~\ref{fig4}). The formation of this small structure may be due to that the convection in the sunspot scar may become different in the condition of the more horizontal field there and further lead to the formation of small loops locally in the sunspot scar. This possible special convection in the sunspot scar may be even the cause for the narrow light bridge that we observe.

\subsection{Possible Link between the Sunspot Scar and the Double-decker Structure}\label{sec25}

\begin{figure*}
\centering
\includegraphics[width=\hsize]{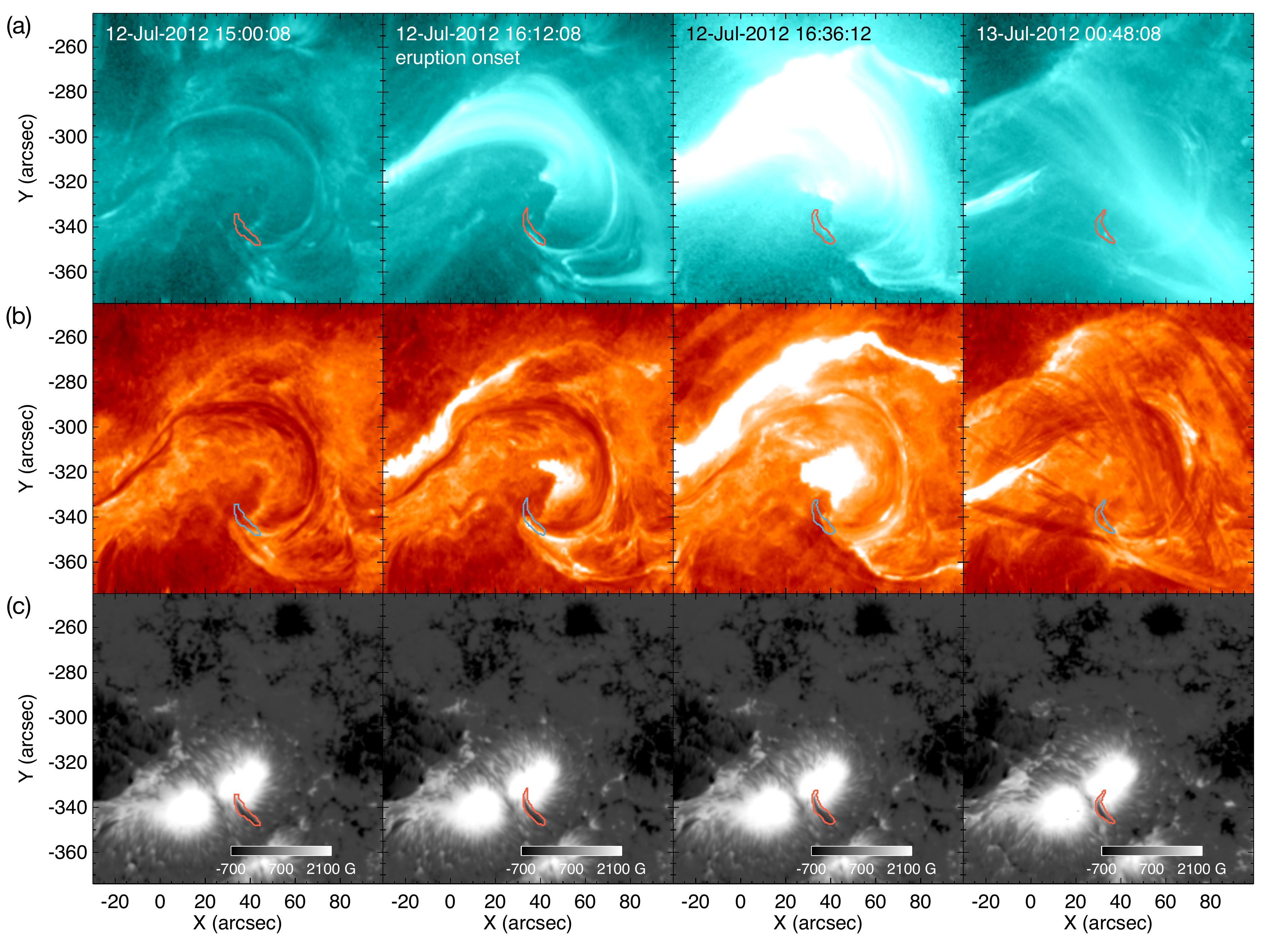}
\caption{131 \AA\ images (panel a), 304 \AA\ images (panel b), and vertical magnetic field images (panel c) of the source region, which are all in helioprojective-cartesian-coordinate and at four moments. The time of images in each column is shown in the first row of them. The red/blue contours in all sub-panels outline the sunspot scar.}
\label{fig6}
\end{figure*}

In Fig.~\ref{fig6}, we show EUV images and vertical magnetic field images in helioprojective-cartesian-coordinates, focusing on the sunspot scar. The red/blue contour, standing for the outline of the sunspot scar, is overlaid on EUV and $B_z$ images. At AIA 131 \AA\ passband, we find that there are flickering bright arcades anchored in the sunspot scar, specifically, at the left boundary of the sunspot scar, before and at the onset of the eruption; actually, the bright arcade at 15:00 UT is the first one mentioned in Section \ref{sec22}. During and after the eruption, due to the overexposure and foreground occlusion, the structures rooted close to the sunspot scar are elusive at 131 \AA\ passband. In addition, as shown at AIA 304 \AA\ passband, the western footpoint of the filament is anchored at the inner side of the sunspot scar at all four moments.

As described in Section \ref{sec22}, the pre-eruptive hot channel appears shortly after the bright arcades appear; it lies above the bright arcades with its two footpoints being always co-spatial with the two ends of the footpoints of the bright arcades in two polarities. This configuration is quite similar to that in the three-dimensional (3D) standard model for the CME/flare \citep{Aulanier2012,Janvier2013,Janvier2014,Aulanier2019}, in which the flux rope field lines are formed by the slipping reconnection. In the model, under the slipping reconnection, the field line anchored at a fixed point in one polarity will evolve from an inclined loop first to a highly sheared field line and finally to a twisted flux rope field line, with its other footpoint slipping along the ribbon in the other polarity (see Fig. 16 in \citeauthor{Dudik2014}\ \citeyear{Dudik2014}). In the present event, the background coronal loops, the bright arcades (those in Fig.~\ref{fig2}a-d as mentioned in Section \ref{sec22}), and the twisted hot channel (see the dashed curve in Fig.~\ref{fig2}d) exactly correspond to the inclined loop, the highly sheared field line, and the twisted flux rope in the model, respectively; the slipping motion along the ribbon in the negative polarity corresponds to that of the soon-to-be flux rope field lines in the model. Therefore, we consider that the bright arcades anchored in the sunspot scar are closely related to the formation of the pre-eruptive hot channel, and even specifically, they represent the highly sheared field lines that will soon evolve into flux rope field lines under the slipping reconnection process. In addition, the filament anchored at the inner side suggests that the field lines rooted there are most likely twisted flux rope field lines. These results strongly suggest that the sunspot scar is closely related to the western footpoint of the double-decker structure.

\section{Sunspot Scars in Observationally-inspired Simulations}
\subsection{Overviews of Modelled CMEs and Sunspot Scars}\label{sec31}
In the following, we further study sunspot scars in two magnetohydrodynamic (MHD) simulations which model the formation of pre-eruptive structures and the eruption of CMEs. We first give brief overviews of these two simulations, and we refer readers to \cite{Zuccarello2015} and \cite{Xing2023} for more details of them. Especially, we note that the parameters in simulations, unless otherwise indicated, are in dimensionless units.

The \textit{Simulation 1} is the ``Run D2'' performed by OHM \citep{Zuccarello2015}, which models the CME eruption by solving zero-$\beta$ MHD equations. The viscosity and resistivity are considered in this simulation. The equations are solved in dimensionless form, and the dimensionless units of the length, time, and magnetic field strength are set to 10 Mm, 67.89 s, and 2.53 G, respectively. It should be mentioned that the set of dimensionless units is for consistency with those of Simulation 2 (see Appendix), but any other choice is possible depending on the event to fit. The initial field of the simulation is a potential dipole, and the dimensionless distance between the centers of two polarities is about 2. Mimicking observations, the initial field is first driven into a highly sheared state by the line-tied shearing flow imposed at the bottom $z=0$ (10 $t_A \le t \le$ 100 $t_A$; $t$ is the dimensionless time, and $t_A$ is the Alfvén time unit). Later, driven by the prescribed line-tied converging flow, a pre-eruptive flux rope of the CME is gradually formed by the magnetic reconnection among sheared arcades, in the framework of the flux-cancellation model (105 $t_A \le t \le$ 164 $t_A$). After $t=164$ $t_A$, the whole system is relaxed by keeping the bottom boundary stationary; meanwhile, as triggered by the torus instability (TI; \citeauthor{Kliem2006} \citeyear{Kliem2006}), the flux rope erupts freely as a CME after $t=164$ $t_A$.

The \textit{Simulation 2} is the ``Simulation Ue'' in \cite{Xing2023}, which studies the CME eruption in a finite-$\beta$ condition (see Appendix \ref{appendix} and \citeauthor{Xing2023}\ \citeyear{Xing2023} for more details about the numerical setups including dimensionless units and also the kinematics of the flux rope). The simulation is performed by MPI-AMRVAC \citep{Xia2018}, and it solves thermal-MHD equations, which include an internal energy equation and incorporate gravity, viscosity, resistivity, and thermal conduction. The initial condition of the simulation is a combination of a bipolar potential field (where the dimensionless distance between the two polarity centers is also about 2) and a hydrostatic corona with a uniform temperature of 1.6 MK. Similar to the first simulation, the initial field is first driven into a highly sheared field under the line-tied shearing flow imposed at the bottom $z=0.015$ ($0\le t\le18$), then a pre-eruptive flux rope of the CME is formed by the flux cancellation ($18 < t \le 60$) under the line-tied converging flow, and finally the system is relaxed and the CME erupts after that ($60 < t \le 71$). Specifically, we note that the modelled flux rope rises almost linearly before $t=68.5$ while exponentially after that, which implies that the onset time of the CME as usually defined in observational analyses (e.g., \citeauthor{McCauley2015}\ \citeyear{McCauley2015}; \citeauthor{Cheng2020}\ \citeyear{Cheng2020}) is $t=68.5$ in this simulation.

\begin{figure*}
\centering
\includegraphics[width=\hsize]{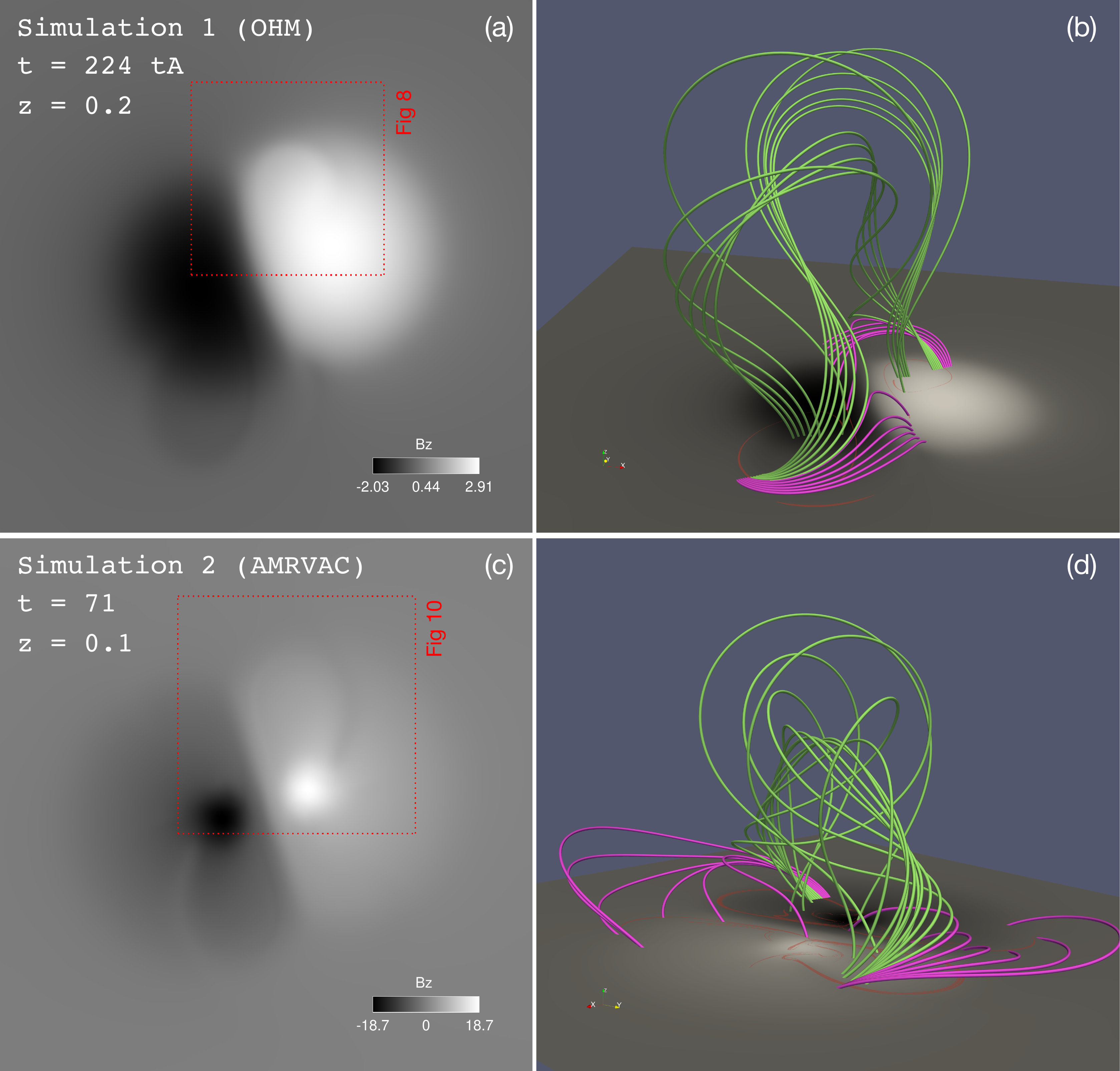}
\caption{Snapshots of sunspot scars and the field lines traced from rakes crossing through sunspot scars for both Simulation 1 and Simulation 2. (a) Snapshot of vertical magnetic field on the plane $z=0.2$ at $t = 224\ t_A$ in Simulation 1. The red dashed box shows the field of view in Fig.~\ref{fig8}. (b) Magnetic field lines traced from two rakes at $t = 224\ t_A$ which cross through the two sunspot scars in the two polarities, respectively. The green tubes represent the CME flux rope field lines and the magenta tubes represent the inclined loops. The bottom plane shows the vertical magnetic field on the plane $z=0.2$, same as that in panel a. The red filled contours show the footprints of QSLs ($\textup{log}Q\ge3$) on this plane. (c) Snapshot of vertical magnetic field on the plane $z=0.1$ at $t = 71$ in Simulation 2. The red dashed box shows the field of view in Fig.~\ref{fig10}. (d) Similar to panel b but for that at $t=71$ in Simulation 2. The bottom plane shows the vertical magnetic field on the plane $z=0.1$, same as that in panel c. The green tubes, magenta tubes, and red filled contours have the same meanings as those in panel b.}
\label{fig7}
\end{figure*}

Aside from the difference in the equations solved, the key differences between the two simulations are the symmetry and the diffusion pattern of the polarities. The dipole in Simulation 1 is asymmetrical while that in Simulation 2 is symmetrical. The converging flow in Simulation 1 diffuses the entire polarities, while the converging flow in Simulation 2 mainly diffuses the periphery of the polarities (see Fig.~\ref{fig7}a,c).

As an overview, in Fig.~\ref{fig7}a, we show the snapshot of vertical magnetic field on the plane $z=0.2$ at $t=224$ $t_A$ for Simulation 1; similarly, we show that on the plane $z=0.1$ at $t=71$ for Simulation 2 in Fig.~\ref{fig7}c. In the following, for Simulations 1 \& 2, we choose to observe the modelled sunspot scars in these two planes which are slightly above the bottom surface, respectively. This choice is due to that the line-tied boundary conditions only allow the vertical magnetic field on the bottom surface to evolve as prescribed by the flows there in the condition of ideal MHD. Since we do not deliberately impose special flows to form the sunspot scar on the bottom surface, no sunspot scar can be present in this line-tied plane. Therefore, we choose an altitude slightly above the bottom: on the one hand, the magnetic field is free to evolve at this altitude under the combined constraints of the underlying line-tying and the overlaying coronal evolution; on the other hand, this altitude above the line-tied plane are sufficiently low to be considered at the bottom of the corona, close enough to the photosphere. We then fined-tuned the choice of these altitudes between the bottom and $z=0.2$ for both two simulations to allow for the best visibility of the scars, within the limits of our models. In addition, this choice means that, when we compare observations with simulations in the following, we implicitly assume that, in observations, the photospheric altitude that corresponds to that of HMI observations is located above the solar depth at which the line-tying is at work.

It is clear that there are sunspot scars in both the positive and the negative polarities for these two simulations, regardless of if the bipolar field is symmetrical or not and how the active region is diffused. For each sunspot scar in each simulation, a bunch of field lines are traced from a rake crossing through it. As shown in panels b and d, some of magnetic field lines are CME flux rope field lines (green tubes) while some others correspond to surrounding inclined loops (magenta tubes). In the following, we will study the magnetic properties of sunspot scars mainly through Simulation 1, considering that its sunspot scars are more similar to that in observations; we will also study the thermal properties of sunspot scars through Simulation 2, benefiting from the thermal-MHD equations it solves.

\subsection{Properties of Modelled Sunspot Scars}\label{sec32}

\begin{figure*}
\centering
\includegraphics[width=\hsize]{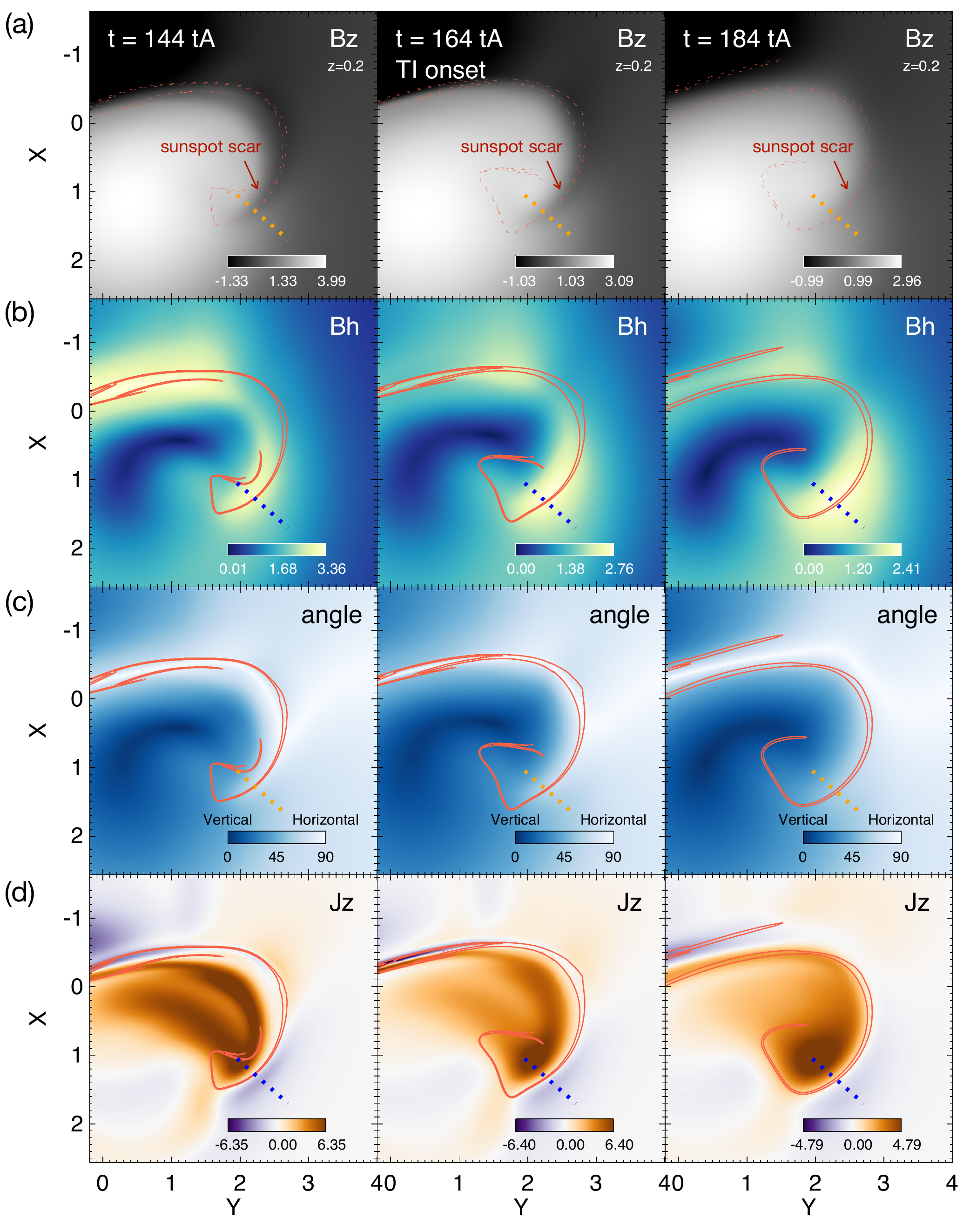}
\caption{Distributions of vertical magnetic field strength ($B_z$; panel a), horizontal magnetic field strength ($B_h$; panel b), inclination angle (angle; panel c), and vertical current density ($J_z$; panel d) on the plane $z=0.2$ at $t=144\ t_A$ (first column), $t=164\ t_A$ (second column), and $t=184\ t_A$ (third column) in Simulation 1. The red dashed or solid contour in each sub-panel shows the outline of the footprint of QSLs in this plane, in which $\textup{log}Q\ge3$. The orange or blue dashed line in each sub-panel marks the slit crossing through the sunspot scar, whose 1D plots are shown in Fig.~\ref{fig9}.}
\label{fig8}
\end{figure*}

\begin{figure*}
\centering
\includegraphics[width=\hsize]{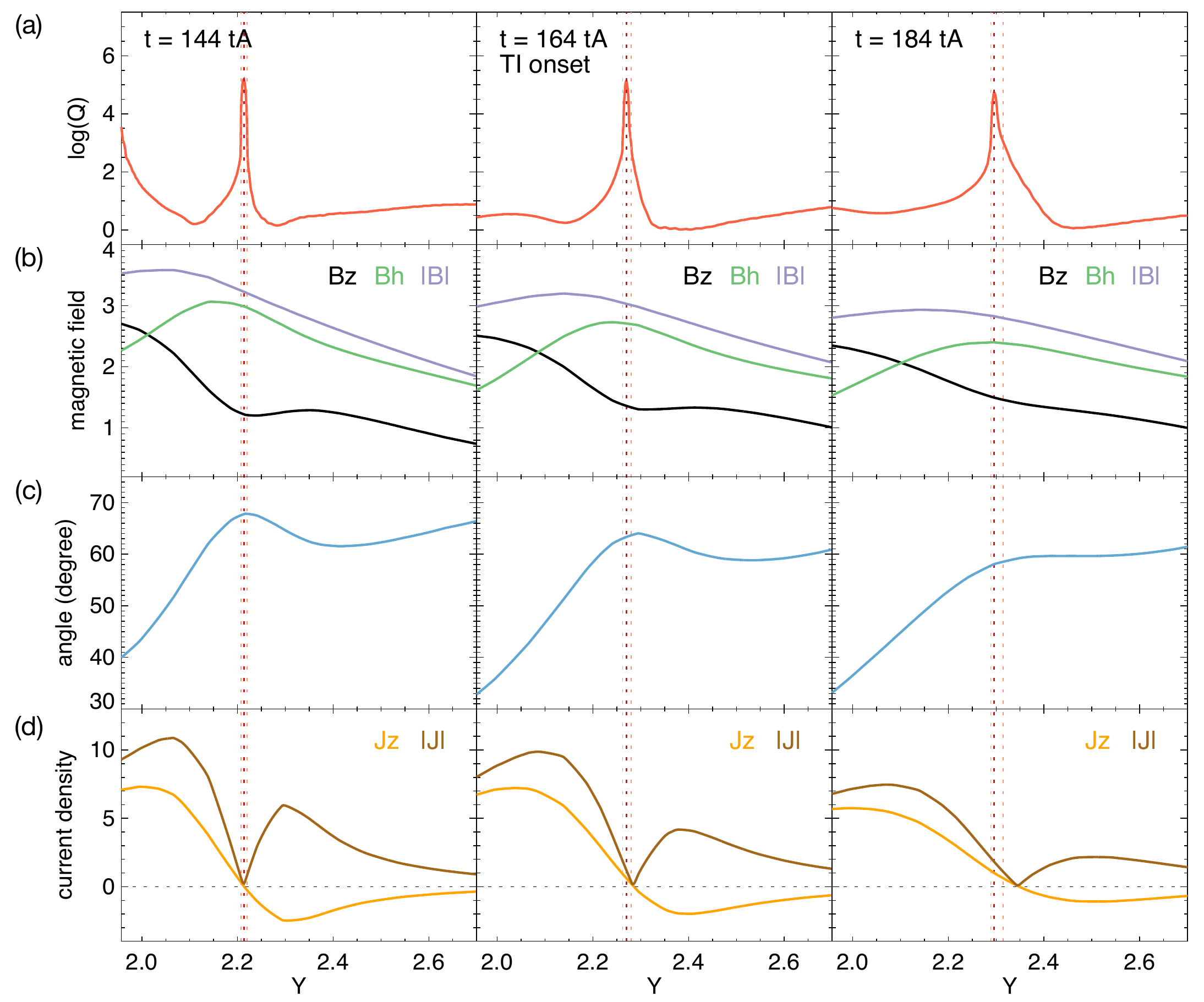}
\caption{The profiles of 1D parameters along the slit (dashed lines in Fig.~\ref{fig8}) crossing through the sunspot scar at $t=144\ t_A$ (first column), $t=164\ t_A$ (second column), and $t=184\ t_A$ (third column) in Simulation 1. From the top to the bottom, they are squashing degree ($\textup{log}Q$; panel a), total magnetic field strength ($|B|$; panel b), horizontal magnetic field strength ($B_h$; panel b), vertical magnetic field strength ($B_z$; panel b), inclination angle (panel c), total current density ($|J|$; panel d), and vertical current density ($J_z$; panel d). The red dashed line in each column marks the peak of $\textup{log}Q$. The pink dashed lines in each column show the range of the QSL footprint ($\textup{log}Q\ge3$). The black horizontal dashed line in panel d marks the value where the current density equals zero.}
\label{fig9}
\end{figure*}

We first study the properties of the modelled sunspot scar in the positive polarity on the plane $z=0.2$ of Simulation 1 at three moments (144 $t_A$, 164 $t_A$ and 184 $t_A$, which are before, at and after the torus instability onset, respectively). As shown in Fig.~\ref{fig8} and Movie 4, the sunspot scar appears and continuously drifts both before and during the eruption, similar with that in observations. We also set a slit crossing through the sunspot scar (marked by the orange and blue dashed lines in Fig.~\ref{fig8}), the detailed properties of the sunspot scar along which are exhibited in Fig.~\ref{fig9}. Obviously, the vertical magnetic field strength is weaker in the sunspot scar (Fig.~\ref{fig8}a and Fig.~\ref{fig9}b). It is also clear that there is a stronger horizontal magnetic field (Fig.~\ref{fig8}b and Fig.~\ref{fig9}b) and a larger inclination angle (Fig.~\ref{fig8}c and Fig.~\ref{fig9}c) in the sunspot scar. The vertical current density is close to zero at the center of the sunspot scar and has a positive peak at the inner side of the sunspot scar and a negative peak at the outer side (Fig.~\ref{fig8}d and Fig.~\ref{fig9}d). In addition, we also find that the total current density is also close to zero at the center of the sunspot scar (Fig.~\ref{fig9}d).

We further study the properties of the sunspot scar in Simulation 2. In Fig.~\ref{fig10}, we show the sunspot scar in the positive polarity on the plane $z=0.1$, and also the detailed information of the sunspot scar along a slit crossing through it (marked by orange/blue dashed lines in Fig.~\ref{fig10}). As shown in Fig.~\ref{fig10} and Movie 5, first, the sunspot scar also has a weaker vertical magnetic field, a stronger horizontal magnetic field, and thus a larger inclination angle; second, the vertical and total current densities are also close to zero at the center of the sunspot scar, and the vertical current density has a positive/negative peak at the inner/outer side. In addition, the temperature at the inner side of the sunspot scar is 0.16 (0.25 MK in dimensional unit) larger than that at the outer side, and the mass density shows a lane of peak at the outer side of the sunspot scar.

We note that the above features of vertical and horizontal magnetic field strength, inclination angle, and vertical current density in both two simulations are in remarkable agreement with the observations, even though these simulations were merely designed for CME onset and certainly not to fit the sunspot scar a priori.

\subsection{Nature of the Sunspot Scar: Flux Rope Footpoint Edge}\label{sec33}
\begin{figure*}
\centering
\includegraphics[width=\hsize]{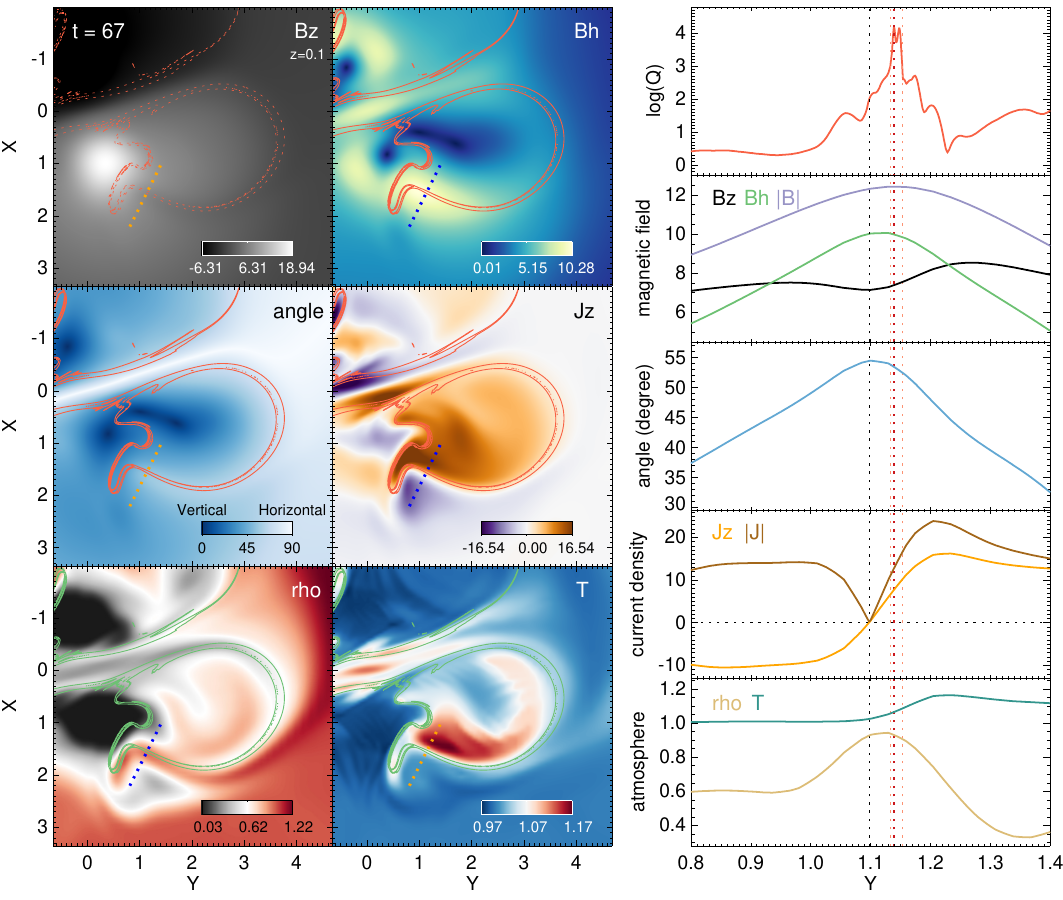}
\caption{Properties of the sunspot scar in the positive polarity of Simulation 2, shown by the 2D images (left half) and 1D plots of slit crossing through the scar (right half). Left half: distributions of vertical magnetic field strength ($B_z$), horizontal magnetic field strength ($B_h$), inclination angle (angle), vertical current density ($J_z$), mass density (rho), and temperature ($T$) on the plane $z=0.1$ at $t=67$ in Simulation 2. The red (or green) solid (or dashed) contours outline the QSL footprints on this plane. Right half: 1D plots along the slit crossing through the sunspot scar at $t=67$ (orange and blue dashed lines in the left half). From the top to the bottom, they are squashing degree ($\textup{log}Q$), total magnetic field strength ($|B|$), horizontal magnetic field strength ($B_h$), vertical magnetic field strength ($B_z$), inclination angle, total current density ($|J|$), vertical current density ($J_z$), temperature ($T$), and mass density (rho). The red dashed line shows the peak of $\textup{log}Q$, and the pink dashed lines demonstrate the range of the QSL footprint ($\textup{log}Q\ge3$). The black vertical dashed line marks the dip of $B_z$ in the sunspot scar. The black horizontal dashed line in the fourth row marks the value where the current density equals zero.}
\label{fig10}
\end{figure*}

To explore the nature of the sunspot scar, we further study the squashing degree $Q$, which measures the mapping of the field lines, on the plane $z=0.2$ for Simulation 1 and on the plane $z=0.1$ for Simulation 2. The squashing degree is calculated by \citep{Titov2002}:
\begin{equation}
Q=\frac{(\frac{\partial X}{\partial x})^2+(\frac{\partial X}{\partial y})^2+(\frac{\partial Y}{\partial x})^2+(\frac{\partial Y}{\partial y})^2}{\vert\frac{\partial X}{\partial x}\frac{\partial Y}{\partial y}-\frac{\partial X}{\partial y}\frac{\partial Y}{\partial x}\vert},
\end{equation}
in which $(x,y)$ and $(X,Y)$ are coordinates of two footpoints of a field line. As exhibited in Fig.~\ref{fig8} and Fig.~\ref{fig10}, the (red/green dashed/solid) contour of $\textup{log}Q=3$ shows the outline of the footprint of QSLs, in which the connectivity of field lines varies sharply in space. In Fig.~\ref{fig9}a and Fig.~\ref{fig10}, we also show $\textup{log}Q$ along the slit crossing through the sunspot scar, for Simulations 1 and 2, respectively. For Simulation 1, Fig.~\ref{fig8} and Fig.~\ref{fig9} clearly show that a section of the hooked part of the QSL footprint matches well and drifts equally with the sunspot scar at all three moments. The similar result for Simulation 2, that the sunspot scar is co-spatial with a section of the hooked part of the QSL footprint, is demonstrated in Fig.~\ref{fig10}. 

\begin{figure*}
\centering
\includegraphics[width=\hsize]{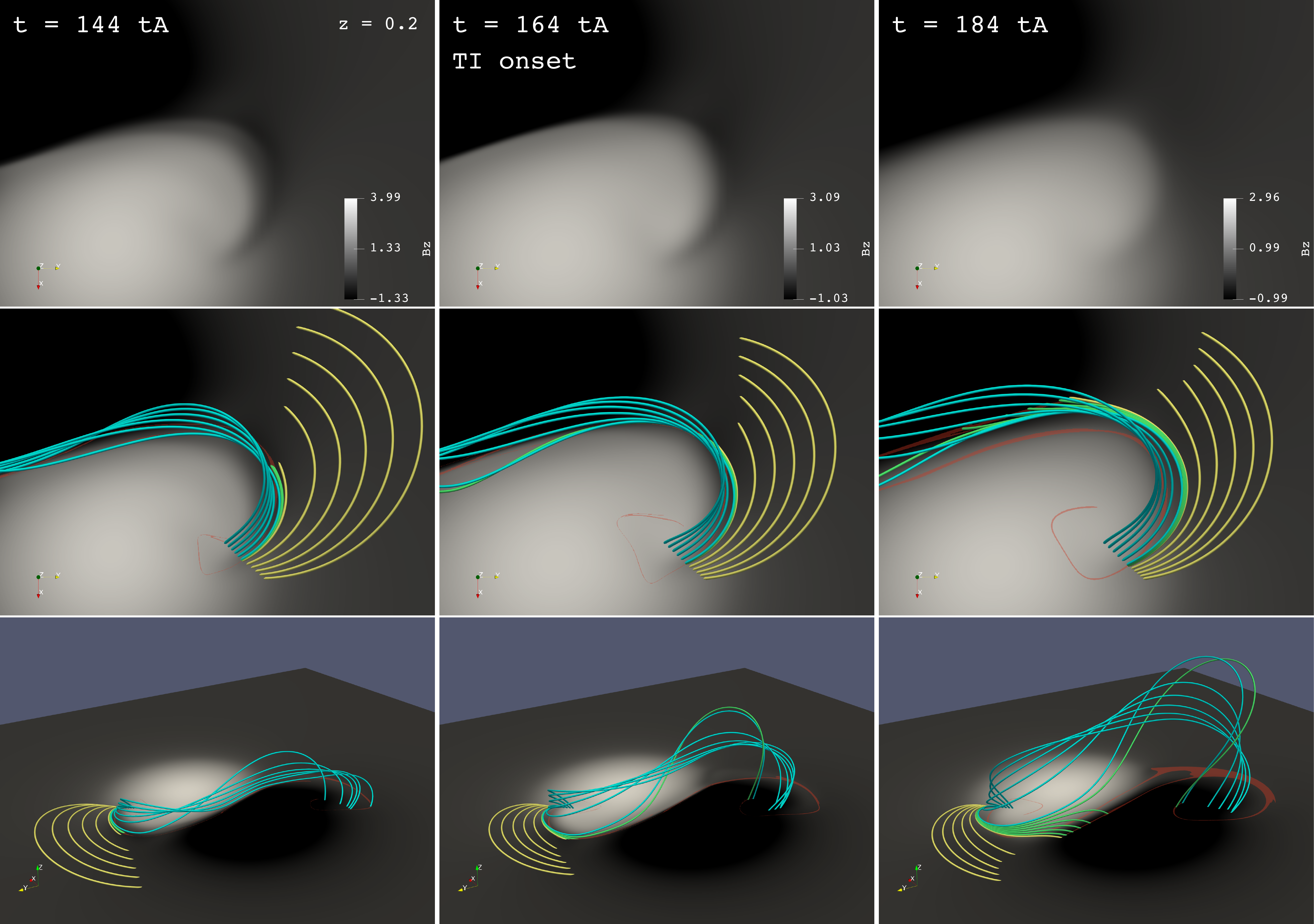}
\caption{Snapshots of sunspot scars and the field lines traced from points along a slit crossing through the sunspot scar in the positive polarity at $t=144\ t_A$ (first column), $t=164\ t_A$ (second column), and $t=184\ t_A$ (third column) in Simulation 1. The bottom plane in each sub-panel shows the vertical magnetic field on the plane $z=0.2$, clearly exhibiting the sunspot scar in the positive polarity. The tubes in the second row represent the field lines traced from points on a slit which is along the slit in Fig.~\ref{fig8}. The blue tubes represent the flux rope field lines anchored at the inner side of the sunspot scar, the yellow tubes represent the inclined loops anchored at the outer side, and the green tubes represent the field lines which are anchored in the scar and at the transition between the previous two bunches of field lines. The red filled contours show the QSLs footprints on the plane $z=0.2$. The third row is the oblique view of the the second row.}
\label{fig11}
\end{figure*}

\begin{figure*}
\centering
\includegraphics[width=\hsize]{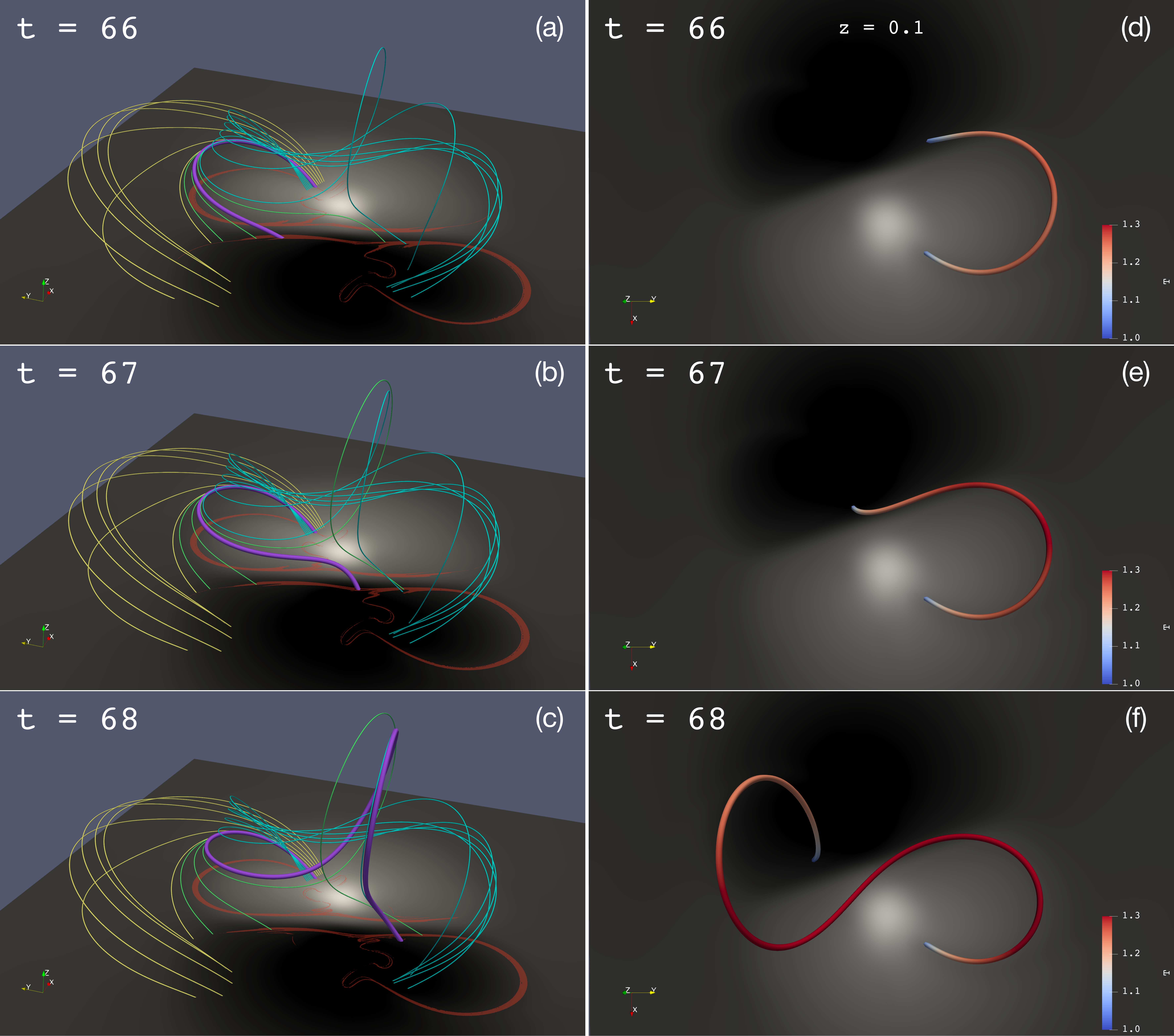}
\caption{Snapshots of field lines traced from points on a slit crossing through the sunspot scar in the positive polarity during $66\le t\le68$ in Simulation 2. The bottom plane in each sub-panel exhibits the vertical magnetic field strength on the plane $z=0.1$. In panels a-c, the thin yellow, green and blue tubes represent the field lines traced from points on a slit which is along the slit in Fig.~\ref{fig10}; the thick purple tube represents the field line which is traced from a fixed point at the bottom plane ($z=0.015$) and crosses through the plane $z=0.1$ with the intersection in the sunspot scar and on the slit. The red filled contours show the QSLs footprints on the plane $z=0.1$. Panels d-f show the top view of panels a-c, with only the plane $z=0.1$ and the thick tube (same as the purple tube in panels a-c) exhibited. The color of the tube shows the temperature along the field line.}
\label{fig12}
\end{figure*}

In addition, in Fig.~\ref{fig11}, we exhibit field lines traced from points along a slit (which is along the slit in Fig.~\ref{fig8}) crossing through the sunspot scar in Simulation 1. It is clear that the field lines traced from points at the inner side of both the QSL footprint and the sunspot scar are flux rope field lines (blue tubes), while those traced from the other side are inclined loops (yellow tubes). The field lines traced from points in the sunspot scar (green tubes), with their other footpoints anchored on the QSL footprint in the negative polarity, mark the transition between the above two bunches of field lines. Similar results for Simulation 2 are shown in Fig.~\ref{fig12}a-c, in which the field lines are also traced from points along a slit which crosses through the scar along the slit in Fig.~\ref{fig10}. The field lines traced from points at the inner side of, at the outer side of, and within the sunspot scar are also flux rope field lines (blue tubes), inclined loops (yellow tubes), and the transition between the previous two (green tubes), respectively.

In short summary, the above results (about QSL footprints and field lines) strongly indicate that the sunspot scar marks the edge of the footpoint of the flux rope both before and during the eruption, in agreement with our implication in observations that the sunspot scar is closely related to the flux rope footpoint of the double-decker structure.

Furthermore, as a thermal-MHD simulation, Simulation 2 unveils more thermal properties of the field lines anchored in the sunspot scar. The tube in Fig.~\ref{fig12}d-f (which is the same as the purple tube in panels a-c) reveals the temporal evolution in thermodynamics of a field line rooted inside the sunspot scar. This field line is traced from a fixed point in the positive polarity at the bottom surface ($z=0.015$) where the velocity is zero during $66\le t\le68$, which ensures that the footpoint of this field line at the fixed point hardly moves. From $t=66$ to $t=68$, this field line crosses through the sunspot scar on the plane $z=0.1$ with the intersection on the slit. In this period, it evolves from an inclined loop to a highly sheared field line and finally to a flux rope field line, with its footpoint in the negative polarity slipping along the QSL footprints (see Fig.~\ref{fig12}a-c). Meanwhile, its temperature increases over time with the maximum temperature along the field line increasing by 0.12 (0.20 MK in dimensional unit) from $t=66$ to $t=68$ (see Fig.~\ref{fig12}d-f). These results indicate that this field line experiences a slipping reconnection and it is heated by the reconnection, when it is anchored in the sunspot scar.

Especially, we note that the evolution of the tube in Fig.~\ref{fig12}d-f is highly consistent with the evolution of the bright arcades/hot channel in observations (see Fig.~\ref{fig2}a-d). Both the modelled field line and the bright arcades in observations are less hot/bright inclined loops at the beginning. Afterwards, they experience the reconnection, with one of their footpoints slipping along the QSL footprint/bright ribbon; meanwhile, they become hotter/brighter as being heated during the reconnection. Finally, they evolve into a flux rope field line/hot channel thread when their footpoints slip to the flux rope footpoint/ribbon hook.

As a summary, we conclude that, on the one hand, the field lines rooted in the sunspot scar spatially represent the transition between the flux rope and the inclined loops at each moment; on the other hand, the temporal evolution of each field line in the sunspot scar shows the transformation from the inclined loops to the flux rope under the slipping reconnection over time.

Finally, it is worth discussing the temperature of the flux rope and related heating processes in Simulation 2. As described in Sections \ref{sec32} and \ref{sec33}, the flux rope (field line) is obviously hotter than the coronal loops, but the temperature difference between these two is not large. This is due to that the maximum magnetic field strength in Simulation 2 is only about 110 G, quite less than that in observations, which leads to insufficient Ohmic heating during the formation of the modelled flux rope. In addition, the setting of uniform resistivity in the simulation may underestimate the resistivity in the reconnection region, which could also result in the flux rope (field line) not being heated as much as that in observations.

\subsection{Explanations of the Sunspot Scar in the Observational Event}\label{sec34}

Considering the high agreement of our simulations with the observations, we explain the relationship between the sunspot scar and the double-decker structure in the observational event on 2012 July 12, with reference to the nature of the sunspot scar as summarized from our simulations.

Before the eruption onset, the sunspot scar drifts towards the south-east and sweeps over the coronal loop regions. In this process, the coronal loops are successively involved in the slipping reconnection, with their footpoints in the positive polarity relatively passing from the outer side of the sunspot scar to its inner side. During the reconnection, these loops are transformed to highly sheared field lines and meanwhile get heated, appearing as the bright arcades anchored in the sunspot scar; they completely evolve into flux rope field lines when they are anchored at the inner side. 

Furthermore, under an extra slipping reconnection, these flux rope field lines slip from the filament footpoint region to the hot channel footpoint region, along a lane connecting the end of the sunspot scar in the main positive sunspot and the adjacent positive sunspot (see Fig.~\ref{fig3}a), after which they appear as loops in the hot channel.

Therefore, we infer that the sunspot scar represents the edge of the footpoint of the pre-eruptive hot channel in observations, which is in agreement with the results in simulations, although the hot channel field lines which are formed in the sunspot scar soon slip to the adjacent sunspot under further reconnection due to the particular complex flux distribution in this non-bipolar active region (see Section \ref{sec23}).

During and after the eruption, it is unclear to determine the relation between the sunspot scar and the eruptive hot channel, due to the foreground occlusion and the further slipping motion along the western faculae. We may suspect that the reconnection is still ongoing in the sunspot scar as the scar still continuously drifts in this period. The low-lying filament-related flux rope may also contribute to the sunspot scar.

Finally, we note that both the hot channel and the filament in observations and the flux ropes in simulations are forward-S with a right-handed chirality. Therefore, the direct current along the flux rope field line should be in the same sign with the magnetic field, that is, positive in the positive polarity; the return current, which shields the flux rope from the background field, should be antiparallel with the magnetic field, that is, negative in the positive polarity (see Table 1 in \citeauthor{Schmieder2018}\ \citeyear{Schmieder2018}). The results in observations and simulations are exactly consistent with this definition: in observations, the vertical current density at the inner/outer side of the sunspot scar which is inside/outside the flux rope footpoint is positive/negative, respectively (see Fig.~\ref{fig4}f and Fig.~\ref{fig5}d); similar results, in agreement with the observations, are also shown in simulations (see Fig.~\ref{fig8}d and Fig.~\ref{fig9}d for Simulation 1 and Fig.~\ref{fig10} for Simulation 2).

\subsection{Mismatches between Observations and Simulations}\label{sec35}
Beyond all of the above agreement, the observations and simulations still reveal a few inconsistencies. First, $B_z$, $B_h$, inclination angle, and $J_z$ in the sunspot scar become more and more inconspicuous over time in Simulation 1 (see Fig.~\ref{fig9}b-d). This implies that the modelled sunspot scar is fading away from before to during the eruption, which does not occur in observations. We infer this difference may be due to that the resistivity in the simulation, which is obviously larger than that in the Sun, may over diffuse the boundary between eruptive vertical field lines and non-eruptive inclined loops and thus both flatten and widen the sunspot scar. The flattening effect indeed smooths the dip/peak of magnetic field and current in the sunspot scar, while the widening effect makes the sunspot scar less visually obvious. This inference is strongly supported by the accelerating fading of the sunspot scar in Simulation 1 after $t=164$ (see Movie 4), since which the resistivity is further increased to ensure the stability of the code (see setups in Section 2.3 in \citeauthor{Zuccarello2015}\ \citeyear{Zuccarello2015}). In addition, we note that, on the one hand, the fading of the sunspot scar directly results from the over-dissipation of the magnetic field in it; on the other hand, the over-dissipation of the magnetic field in regions above the sunspot scar also affects the sunspot scar indirectly by Alfvén waves. Lastly, this mismatch might also be due to that the surviving low-lying filament-related flux rope may also contribute to the sunspot scar in the observation while the low-lying structure (see Fig. 2a in \citeauthor{Aulanier2019}\ \citeyear{Aulanier2019}) does not play such a role in the simulation.

Second, the total magnetic filed strength does not show a dip in the modelled sunspot scars at any moments (Fig.~\ref{fig9}b and Fig.~\ref{fig10}), which is different from that in observations. We speculate that it is also the excessive resistive dissipation in the simulation that erases this feature, especially considering the total magnetic field strength is only reduced a bit in the observational sunspot scar.

Especially, taking the sunspot scar at $t=144\ t_A$ in Simulation 1 as an example, although $B_z$ at its center is decreased by about 30-40\%, its reduction of $B_z$ is still less visually obvious than that at 14:58 UT in observations (see first column of Fig. \ref{fig5}b; $B_z$ is also decreased by 39\% at the center of the scar). We point out that this visual effect may be due to the widening of the modelled sunspot scar. The wide dip of $B_z$ in the simulation may result from not only the over-dissipation mentioned above, but also that the spatial resolution of the simulation limits the minimum width of the modelled sunspot scar.

\section{Summary}
\begin{figure*}
\centering
\includegraphics[width=\hsize]{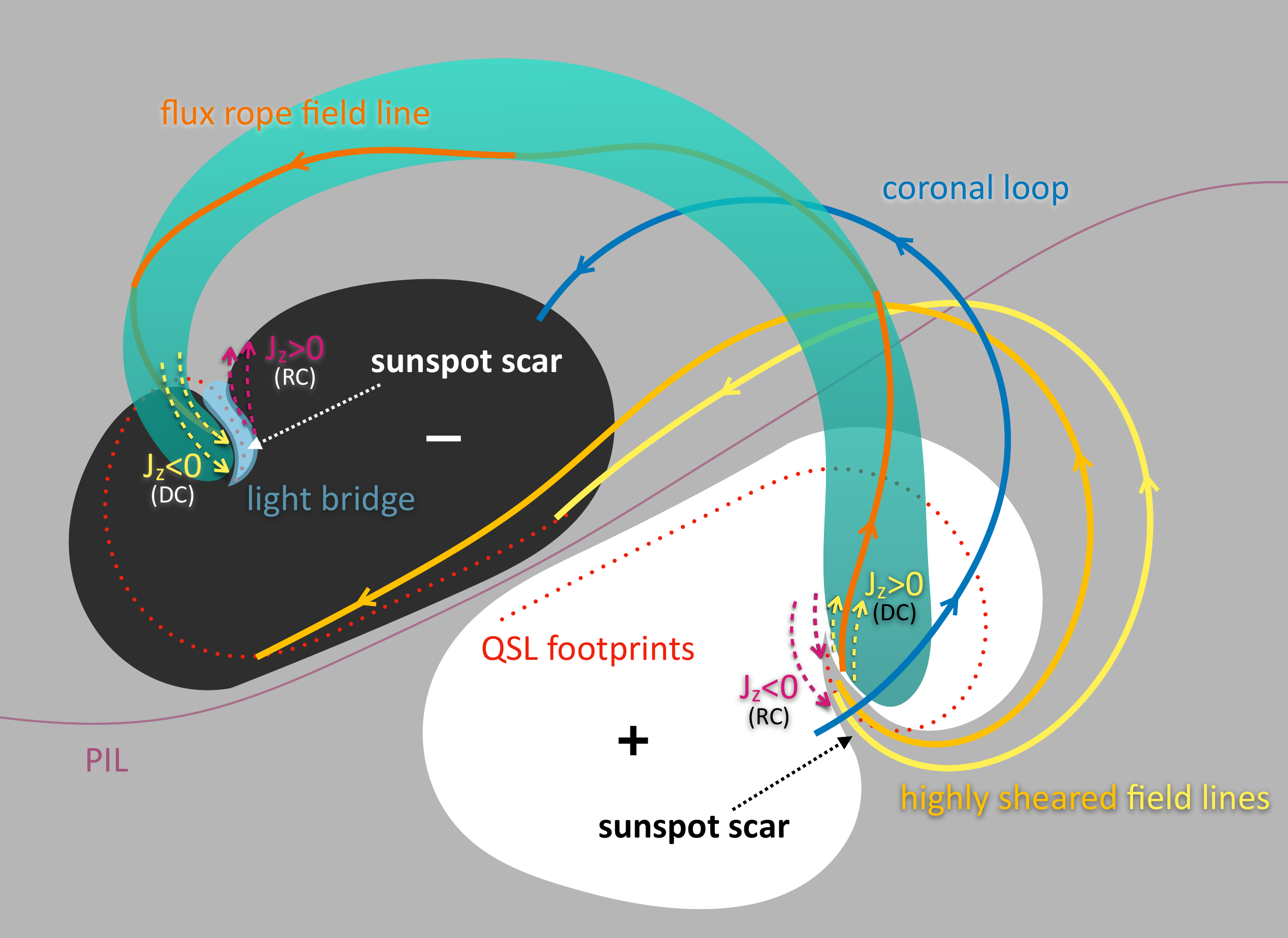}
\caption{A sketch of the sunspot scar. The white and black regions represent the positive and negative polarities in an active region, respectively, which are separated by the PIL (the purple solid curve). The sunspot scar is shown as the arc-shaped intrusion in the positive/negative polarity. We only show the light bridge which is co-spatial with the sunspot scar in the negative polarity, while the light bridge could also appear in the positive one although not shown. In the positive polarity, we show the flux rope (represented by the green tube and also the dark orange field line in it) anchored at the inner side of the sunspot scar, the coronal loop (blue field line) rooted at the outer side, and the highly sheared field lines (light orange and yellow field lines) whose one footpoint is in the sunspot scar and the other one is on the QSL footprints (shown by the red dashed curves) in the other polarity. Similarly, the above field lines also exist around the sunspot scar in the negative polarity although not shown. The yellow and purple dashed arrows at two sides of the sunspot scars represent the direct current (DC) in the forward-S flux rope and the return current (RC) in the shielding background coronal loops, respectively.}
\label{fig13}
\end{figure*}

The properties and the nature of the sunspot scar are summarized as a sketch in Fig.~\ref{fig13}:
\begin{enumerate}
\item Before and during the flux rope eruption, the sunspot scar manifests as an arc-shaped structure intruding in the positive/negative polarity and appears as a light bridge in the white light passband. The magnetic field in the sunspot scar is more inclined with a stronger horizontal magnetic field and a weaker vertical magnetic field relative to those at the surrounding umbra. In addition, the sunspot scar is more homogeneous than the surrounding penumbra, in the latter of which the distributions of inclination angle and $B_z$ are intercombed. Both the vertical current density and the total current density are close to zero at the center of the sunspot scar, while they both have two peaks at two sides.
\item Spatially, the sunspot scar represents the edge of the flux rope footpoint. The sunspot scar matches well with a section of the hooked part of QSL footprints. At the inner side of the sunspot scar, the magnetic field lines are flux rope field lines and carry parallel direct current (DC) in condition of a forward-S flux rope; while, at the outer side, the field lines are coronal loops where the return current (RC) is antiparallel to the magnetic field in the same condition. The field lines rooted in the sunspot scar are highly sheared, as a spatial transition between the previous two bunches of field lines.
\item Temporally, the sunspot scar continuously drifts before and during the eruption, showing the drifting and the deformation of the flux rope footpoint \citep{Aulanier2019,Lorincik2019,Xing2020,Gou2023}. As an area that was originally at the outer side is swept by the sunspot scar and is brought to the inner side, the field lines anchored there experience a slipping reconnection with their footpoints in the other polarity slipping along the QSL footprint; they evolve from the background coronal loops to highly sheared field lines and finally to the flux rope field lines and are meanwhile heated by the reconnection.
\end{enumerate}

\section{Discussions}
The light bridge that we study has many characteristics similar to those in the previous studies. For example, \cite{Rueedi1995} and \cite{Leka1997} showed that the magnetic field in the light bridge is more inclined and the total magnetic field strength there is weaker relative to those at the surrounding umbra, which are exactly what we find here. The previous studies generally interpret the light bridge as the field-free (or weak-field) material intruding into the umbra by upward convection; the surrounding umbral fields gather above and cover the intrusive material in a shape of canopy \citep{Leka1997,Toriumi2015}. However, here we show that this specific narrow and curved light bridge is more closely related to the highly inclined/sheared field lines which are in the process of slipping reconnection and at the transition between the flux rope field lines and the background coronal loops. This result, which is obviously different from the previous explanation, provides a new perspective for understanding the nature of at least some light bridges.

Our results also have implications for understanding the heating in the pre-eruptive structure. As demonstrated in the observation \citep{Cheng2023} and the simulation \citep{Xing2023}, the magnetic reconnection in current sheets under (and sometimes also surrounding) the pre-eruptive flux rope could heat the local plasma while forming twisted flux rope field lines. These newly-formed hot field lines are added into the pre-eruptive flux rope and thus contribute to its heating. Here, also with both the observation and the simulation, we show that magnetic field lines are actually gradually heated meanwhile they evolve from coronal loops to flux rope field lines, and especially, both of these two processes are closely related to the slipping reconnection. This indicates that no matter for the entire flux rope or for a certain field line in it, the heating by reconnection is a gradual rather than instantaneous process, which enriches our understandings in \cite{Cheng2023} and \cite{Xing2023}.

As mentioned in Section \ref{sec1}, there are two methods to identify the flux rope footpoint, that, one with the flare ribbon hook and the other with the core dimming. However, both of these two methods face many difficulties in their applications to observations. First, for the flare ribbon hook method, the hook in the observation is usually partially closed, and thus the boundary of the flux rope footpoint at the unclosed section of the hook is elusive to determine with this method.

Second, for the core dimming method, the dimming regions before the eruption \citep{Qiu2017,Wang2019,Wang2023} and during the eruption \citep{Qiu2007,Dissauer2018} are usually determined by empirical thresholds for the emission intensity. The uncertainty of the empirical threshold and the use of a constant threshold during the whole evolution of an event may bring errors to the identification of the footpoint. Especially, the footpoints before the eruption determined in these works are usually fixed. However, the real footpoint should evolve and move with the evolution of the pre-eruptive flux rope (as the observational event in this work). In addition, due to lack of flare ribbon hooks before the eruption, it is no longer possible to limit the empirical threshold of the pre-flare dimming by the ribbon hook as what \cite{Xing2020} did for the dimming region during the eruption.

We suggest that the sunspot scar could alleviate the problems above. For both the pre-eruptive and eruptive flux ropes, the sunspot scar is expected to help (a) directly show the footpoint edge and (b) indirectly determine the footpoint by limiting the threshold/area of the core dimming region, the latter of which is previously only available with clear ribbon hooks during the eruption (e.g., \citeauthor{Wang2017}\ \citeyear{Wang2017}; \citeauthor{Xing2020}\ \citeyear{Xing2020}).

It should be pointed out that the observation and the simulations studied in this work are all in the framework of ``pre-eruptive structures being flux ropes''. Therefore, what we have revealed, that the sunspot scar before the eruption represents the edge of the pre-eruptive flux rope footpoint and that the drifting of the former represents the growth and deformation of the pre-eruptive flux rope, are also within this framework. However, the pre-eruptive structure is also suggested to be a sheared arcade system for some CME events \citep{Antiochos1994,DeVore2000,Song2014,Ouyang2015}. Whether the sunspot scar also exists in the pre-eruptive stage of events in which the pre-eruptive structures are sheared arcades, and if so, what the nature of this type of sunspot scar is, need to be further studied in future works.

We note that sunspot scars are not always observed even in CME/flare events with flux ropes. A reasonable speculation to explain such a rarity is: although the decrease of $B_z$ and also other features at the edge of the flux rope footpoint may exist in many events, sunspot scars could be not visible (1) when their features are not as strong as those in the event on 2012 July 12 or (2) when their features are covered up by the features of surrounding magnetic field when the edge of the flux rope footpoint is located in faculae or penumbra rather than umbrae. This speculation needs to be examined by more investigations on the conditions of formation and observability of sunspot scars in the future.

However, it should also be mentioned that the sunspot scar is not a unique structure existing only in the event on 2012 July 12. For example, although not as clear as the event that we studied, there is also a ``sunspot-scar-like'' structure in the active region NOAA 11818, which is the source region of a CME/flare event on 2013 August 17 (see Fig. 1 and Fig. 6 in \citeauthor{Zhang2022}\ \citeyear{Zhang2022}). It is thus expected that more events with sunspot scars could be discovered in the future, with our work as a starting point. On the basis of more events, the sunspot scar may offer a new opportunity for the ground-based telescopes (such as DKIST, WeHoST, and future EST) to study the properties and evolution of flux ropes related to CMEs, as the sunspot scar and its related light bridge are observable with photosphere-polarimetry instruments in general and with ground-based instruments in particular.

\begin{acknowledgements}
C.X. and G.A. acknowledge financial support from the French national space agency (CNES), as well as from the Programme National Soleil Terre (PNST) of the CNRS/INSU also co-funded by CNES and CEA.
C.X. also acknowledges financial support from China Scholarship Council.
G.A. thanks the Croom team for fruitful discussions.
X.C. and M.D.D. are supported by National Key R\&D Program of China under grants 2021YFA1600504 and NSFC under grant 12127901. 
Solar Dynamics Observatory (SDO) is a mission of NASA's Living With a Star Program.
Simulations analysed in this work were performed on the HPC center MesoPSL which is financed by the Région Île-de-France and the project Equip@Meso of the PIA supervised by the ANR, the computing facilities in High Performance Computing Center of Nanjing University, and the Cholesky computing cluster from the IDCS mesocentre at Ecole Polytechnique.
\end{acknowledgements}

\newpage
\clearpage

\newpage
\begin{appendix}
\section{Details of \textit{Simulation 2}}\label{appendix}
In the following, we introduce the setups of \textit{Simulation 2} and also the kinematics of the flux rope in this simulation. We refer readers to the ``Simulation Ue'' in \cite{Xing2023} for more details of this simulation.

\subsection{Equations and Grids}
The Simulation 2 is performed by the code MPI-AMRVAC \citep{Xia2018} with solving the following equations:
\begin{equation}
\frac{\partial\rho}{\partial t}+\nabla\cdot(\rho\boldsymbol{v}) = 0
\end{equation}
\begin{equation}
\frac{\partial(\rho\boldsymbol{v})}{\partial t}+\nabla\cdot[\rho\boldsymbol{v}\boldsymbol{v}+(p+\frac{\boldsymbol{B}^2}{2\mu_0})\boldsymbol{I}-\frac{\boldsymbol{B}\boldsymbol{B}}{\mu_0}] = \rho\boldsymbol{g}+2\mu\nabla\cdot[\boldsymbol{S}-\frac{1}{3}(\nabla\cdot\boldsymbol{v})\boldsymbol{I}]
\end{equation}
\begin{equation}\label{eq_induction}
\frac{\partial \boldsymbol{B}}{\partial t}+\nabla\cdot(\boldsymbol{v}\boldsymbol{B}-\boldsymbol{B}\boldsymbol{v}+\psi\boldsymbol{I}) = -\nabla\times(\eta\boldsymbol{J})
\end{equation}
\begin{equation}
\frac{\partial e_{\textup{int}}}{\partial t}+\nabla\cdot(e_{\textup{int}}\boldsymbol{v})=-p\nabla\cdot\boldsymbol{v}+2\mu[\boldsymbol{S}:\boldsymbol{S}-\frac{1}{3}(\nabla\cdot\boldsymbol{v})^2]+\eta J^2+\nabla\cdot[\kappa_{||}(\boldsymbol{b}\cdot\nabla T)\boldsymbol{b}]
\end{equation}
\begin{equation}
\nabla\times\boldsymbol{B} = \mu_0\boldsymbol{J}
\end{equation}
\begin{equation}\label{eq_psi}
\frac{\partial\psi}{\partial t}+c_h^2\nabla\cdot\boldsymbol{B} = -\frac{c_h^2}{c_p^2}\psi.
\end{equation}

The equations are solved in the dimensionless form. The magnetic permeability $\mu_0$ is set to 1. The dimensionless unit of the length, time ($t$), mass density ($\rho$), thermal pressure ($p$), internal energy ($e_{\textup{int}}$), temperature ($T$), velocity ($\boldsymbol{v}$) magnitude and magnetic field ($\boldsymbol{B}$) strength is 10 Mm, 67.89 s, $2.34\times10^{-15}$ g cm$^{-3}$, 0.51 erg cm$^{-3}$, 0.51 erg cm$^{-3}$, 1.6 MK, 147.30 km s$^{-1}$ and 2.53 G, respectively. $\boldsymbol{J}$ is the current density. $\boldsymbol{g}=-g\boldsymbol{e}_z$ represents the gravity acceleration. $\mu$, the dynamic viscosity coefficient, is set to $10^{-4}$ in dimensionless unit, and $\eta$ is the resistivity coefficient. $\kappa_{||}$ is the parallel conductivity coefficient and equals to $8\times 10^{-7}T^{5/2}$ erg cm$^{-1}$ s$^{-1}$ K$^{-1}$. $\boldsymbol{S}$ is the strain tensor and $\boldsymbol{I}$ is the unit tensor. $\boldsymbol{b}=\boldsymbol{B}/B$ is the normalized magnetic field. The zero divergence condition of the magnetic field is maintained with the GLM method \citep{Dedner2002} by introducing a parameter $\psi$, the evolution of which follows Equation \ref{eq_psi}.

The physical domain of the simulation, in the range of $-70\ \textup{Mm}\le x\le70\ \textup{Mm}$, $-70\ \textup{Mm}\le y\le70\ \textup{Mm}$, and $0\ \textup{Mm}\le z\le140\ \textup{Mm}$, is resolved by a stretched grid $nx\times ny\times nz=144\times144\times96$. The finest spatial resolutions in three directions are all about 300 km.

\subsection{Initial Conditions and Boundary Conditions}
The initial magnetic field is set to a potential bipolar field:
\begin{equation}
\begin{gathered}
B_x(t=0)=\Sigma_{m=1}^4c_m(x-x_m)r_m^{-3} \\
B_y(t=0)=\Sigma_{m=1}^4c_m(y-y_m)r_m^{-3} \\
B_z(t=0)=\Sigma_{m=1}^4c_m(z-z_m)r_m^{-3} \\
r_m=\sqrt{(x-x_m)^2+(y-y_m)^2+(z-z_m)^2},
\end{gathered}
\end{equation}
where ($c_1=60, x_1=0.9, y_1=0.3, z_1=-1.1$), ($c_2=-60, x_2=-0.9, y_2=-0.3, z_2=-1.1$), ($c_3=45, x_3=9, y_3=3, z_3=-11$), and ($c_4=-45, x_4=-9, y_4=-3, z_4=-11$) in dimensionless units.

The initial atmosphere is set to a hydrostatic corona with a uniform temperature of 1.6 MK. The initial mass density on the plane $z=0$, which corresponds to 3 Mm above the solar surface, is set to $2.34\times10^{-15}$ g cm$^{-3}$. The initial velocity and parameter $\psi$ are set to zero.

We set line-tied boundary conditions for the bottom boundary, open boundary conditions for the top boundary, and closed boundary conditions for the four side boundaries. Here the line-tied conditions mean that the footpoints of field lines can only move horizontally following the prescribed motions in the condition of ideal MHD \citep{Aulanier2005}. We refer readers to \cite{Xing2023} for more details of boundary conditions.

\subsection{Driving Motions}
In Simulation 2, to drive the magnetic field and form a flux rope, two types of line-tied motions are imposed at the \textit{cell-center bottom surface} which is a horizontal surface ($z=0.015$) at the altitude of the cell center of the layer $k=1$ (cell layers in physical domain indexed by $k=1, 2, 3, ..., 96$ from the bottom to the top). According to types of driving motions, the simulation is divided into three phases: the shearing phase ($0\le t\le18$) with a shearing motion ($v_x^s, v_y^s, v_z^s$) applied to drive the initial potential field to a highly sheared state:
\begin{equation}\label{shear}
\begin{gathered}
v_x^s(k=1; t)=\gamma(t)V_x^s(t) \\
v_y^s(k=1; t)=\gamma(t)V_y^s(t) \\
v_z^s(k=1; t)=0 \\
V_x^s(t)=0.16\Psi_0(t)\partial_y\Psi(t) \\
V_y^s(t)=-0.16\Psi_0(t)\partial_x\Psi(t) \\
\gamma(t)=\left\{
\begin{aligned}
\frac{1}{2}\tanh[3.75(t-1)]+\frac{1}{2} & & & & & & 0\le t<16 \\
-\frac{1}{2}\tanh[3.75(t-17)]+\frac{1}{2} & & & & & & 16\le t\le18 \\
\end{aligned}
\right. \\
\Psi(t)=\exp[-5.5(\frac{B_z(k=1; t)}{B_z^{max}(k=1; t)})^2],
\end{gathered}
\end{equation}
the converging phase ($18< t\le60$) with a converging motion ($v_x^c, v_y^c, v_z^c$) applied to form a flux rope by the flux cancellation:
\begin{equation}\label{converge}
\begin{gathered}
v_x^c(k=1; t)=\gamma(t)V_x^c(t) \\
v_y^c(k=1; t)=\gamma(t)V_y^c(t) \\
v_z^c(k=1; t)=0 \\
V_x^c(t)=0.16\Psi_0(t)\partial_x\Psi(t) \\
V_y^c(t)=0.16\Psi_0(t)\partial_y\Psi(t) \\
\gamma(t)=\left\{
\begin{aligned}
\frac{1}{2}\tanh[3.75(t-19)]+\frac{1}{2} & & & & & & 18<t\le59 \\
-\frac{1}{2}\tanh[6.0(t-59.5)]+\frac{1}{2} & & & & & & 59<t\le60 \\
\end{aligned}
\right. \\
\Psi(t)=\exp[-27.5(\frac{B_z(k=1; t)}{B_z^{max}(k=1; t)})^2],
\end{gathered}
\end{equation}
and the relaxation phase $(60<t\le71)$ in which the velocity in the layer $k=1$ is fixed to zero to relax the whole system. It should be noted that Equations \ref{shear} and \ref{converge} are in dimensionless form.

Correspondingly, in the layer $k=1$, the dissipation term in Equation \ref{eq_induction} is modified into:
\begin{equation}
\frac{\partial \boldsymbol{B}}{\partial t}+\nabla\cdot(\boldsymbol{v}\boldsymbol{B}-\boldsymbol{B}\boldsymbol{v}+\psi\boldsymbol{I}) = \left\{
\begin{aligned}
0 & & & & & 0\le t\le18 \\
\eta(\frac{\partial^2}{\partial x^2}+\frac{\partial^2}{\partial y^2})\boldsymbol{B} & & & & & 18<t\le60. \\
0 & & & & & 60< t\le71 \\
\end{aligned}
\right.
\end{equation}
During the shearing phase and relaxation phase, there is no dissipation term in the layer $k=1$ to fulfill the line-tied condition; the dimensionless resistivity is set to $10^{-4}$ in the whole region except the layer $k=1$. During the converging phase, a two-dimensional (2D) dissipation term \citep{Aulanier2010} is set in the layer $k=1$ to allow the flux cancellation close to the PIL; the dimensionless resistivity is set to $4\times10^{-4}$ in the whole region including the layer $k=1$.

\subsection{Kinematics of the Flux Rope}
The kinematics of the flux rope in Simulation 2 is estimated by measuring the height of the apex of an overlying field line right above the flux rope. The overlying field line is traced from a point at the center of the positive polarity and on the cell-center bottom surface. The velocity at this point is quite small or even zero in all phases, which ensures that this field line could well reflect the kinematics of the flux rope.

\begin{figure*}
\centering
\includegraphics[width=0.4\hsize]{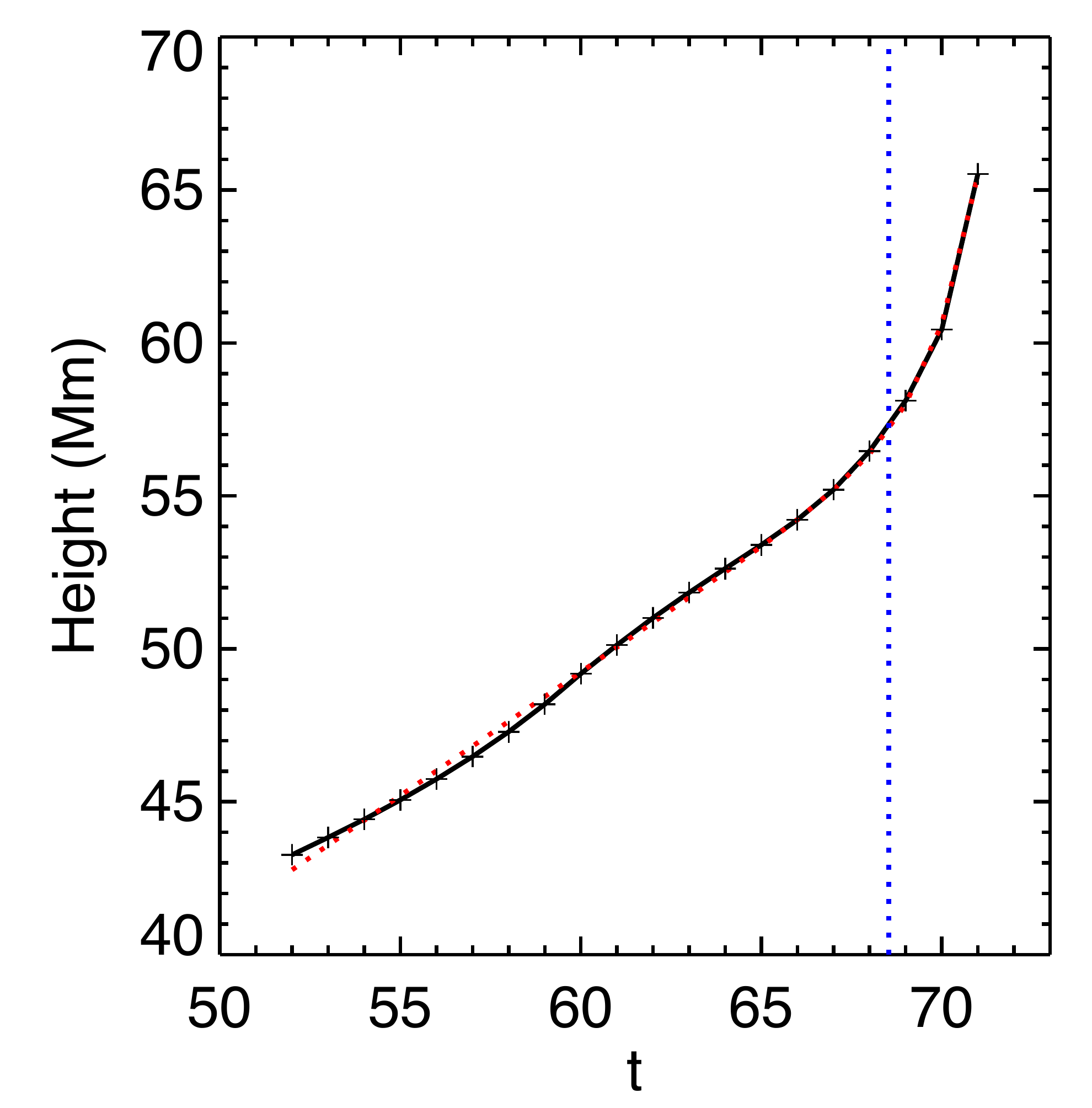}
\caption{Kinematics of the flux rope in Simulation 2. The black points and the black curve represent the evolution of the measured height of the (pre-)eruptive structure, and the red dashed curve is the best fitting curve of the measured height-time data. The vertical dashed line marks the start time of the exponential rise of the CME.}
\label{figs1}
\end{figure*}

The height evolution of the flux rope before and during the eruption $(52\le t\le71)$ is shown in Fig. \ref{figs1}. The onset time of the CME in the observational sense, that is, the start time of the exponential rise of the CME, is determined with methods in \cite{McCauley2015} and \cite{Cheng2020}. The evolution of the height ($h$) of the flux rope with the time ($t$) in Fig. \ref{figs1} is fitted by a function $h=ae^{bt}+ct+d$, which is composed of an exponential term $h_e=ae^{bt}$ and a linear term $h_l=ct+d$. The start time of the exponential rise is defined at $t=\textup{ln}(c/ab)/b$, after which the velocity contributed by the exponential term, $v_e=abe^{bt}$, exceeds that contributed by the linear term, $v_l=c$. For Simulation 2, this start time is at $t=68.5$, marked by the blue dashed line in Fig. \ref{figs1}.

\end{appendix}

\end{document}